%% file: TCAD_LLMAnalog.tex
\documentclass[10pt, journal]{IEEEtran}
\input{./texlib/ieee_setting.tex}

\graphicspath{{./figs/}}

\IEEEoverridecommandlockouts

\usepackage{ulem}
\input{./texlib/ieee_setting}
\graphicspath{{./figs/}}

\usepackage{ulem}
\usepackage{color}
\usepackage{xcolor}
\usepackage{pifont}
\usepackage{makecell}
\usepackage{markdown}
\usepackage{tcolorbox}
\usepackage{amsmath}
\usepackage{graphicx}
\usepackage{hyperref}

\title{
\textbf{
LayoutCopilot: An LLM-powered Multi-agent Collaborative Framework for Interactive Analog Layout Design
}
}

\author{
    
    Bingyang Liu$^\dagger$,
    Haoyi Zhang$^\dagger$,
    Xiaohan Gao, 
    Zichen Kong~\IEEEmembership{Student Member,~IEEE}, \\
    Xiyuan Tang~\IEEEmembership{Senior Member,~IEEE}, 
    Yibo Lin~\IEEEmembership{Member,~IEEE}, 
    Runsheng Wang~\IEEEmembership{Member,~IEEE}, \\
    Ru Huang~\IEEEmembership{Fellow,~IEEE}

    \thanks{
    This work is supported in part by the National Science Foundation of China (Grant No. 62141404, 62125401, 62034007), the Natural Science Foundation of Beijing, China (Grant No. Z230002), STIC (Grant No. QYJS-2023-2303-B), and the 111 project (B18001).
    }
    
    \thanks{$^\dagger$ Equal contribution.}
    \thanks{B.~Liu is with the School of Electronics Engineering and Computer Science, Peking University, Beijing, China.}
    \thanks{H.~Zhang and Z.~Kong are with the School of Integrated Circuits, Peking University, Beijing, China.}
    \thanks{X.~Gao is with the School of Computer Science and the School of Integrated Circuits, Peking University, Beijing, China.}
    \thanks{X.~Tang is with the Institute for Artificial Intelligence and the School of Integrated Circuits at Peking University, and the Beijing Advanced Innovation Center for Integrated Circuits, Beijing, China.}
    \thanks{Y.~Lin, R.~Wang, and R.~Huang are with the School of Integrated Circuits, Peking University, Beijing Advanced Innovation Center for Integrated Circuits, Beijing, China. and Institute of Electronic Design Automation, Peking University, Wuxi, China}
    
    \thanks{Corresponding authors: Yibo Lin (\url{yibolin@pku.edu.cn})}
    
}

\newcommand{\chg}[1]{\textcolor{black}{#1}}
\newcommand{\scdchg}[1]{\textcolor{black}{#1}}

\begin{document}


\maketitle

\input{doc/abstract}


\input{doc/intro}
\input{doc/prelim}
\input{doc/algo}

\input{doc/result}

\input{doc/conclu}

{
\small
\bibliographystyle{IEEEtran}



}

\input{doc/bio}

\end{document}

%% file: texlib/ieee_setting.tex
\usepackage{algpseudocode}    
\usepackage{algorithm}
\usepackage{amsmath}
\usepackage{amssymb}
\usepackage{amsfonts}
\usepackage{booktabs}
\usepackage{multirow}
\usepackage{subfigure}
\usepackage{graphicx}         
\usepackage{color}
\usepackage{cite}             
\usepackage{comment}          
\usepackage{soul}             
\soulregister\cite7
\soulregister\ref7
\soulregister\pageref7
\usepackage{amsthm}
\usepackage{etoolbox}         
\usepackage{url}
\usepackage{nth}              
\usepackage{bm}               

\makeatletter
\let\OldStatex\Statex
\renewcommand{\Statex}[1][3]{%
  \setlength\@tempdima{\algorithmicindent}%
  \OldStatex\hskip\dimexpr#1\@tempdima\relax
}
\makeatother

%% file: doc/abstract.tex
\begin{abstract}

    Analog layout design heavily involves interactive processes between humans and design tools. Electronic Design Automation (EDA) tools for this task are usually designed to use scripting commands or visualized buttons for manipulation, especially for interactive automation functionalities, which have a steep learning curve and cumbersome user experience, making a notable barrier to designers' adoption. Aiming to address such a usability issue, this paper introduces LayoutCopilot, a pioneering multi-agent collaborative framework powered by Large Language Models (LLMs) for interactive analog layout design. LayoutCopilot simplifies human-tool interaction by converting natural language instructions into executable script commands, and it interprets high-level design intents into actionable suggestions, significantly streamlining the design process. Experimental results demonstrate the flexibility, efficiency, and accessibility of LayoutCopilot in handling real-world analog designs.
\end{abstract}

%% file: doc/intro.tex
\section{Introduction}
\label{sec:Introduction}

\begin{figure}[tb]
    \centering
    \includegraphics[width=0.48\textwidth]{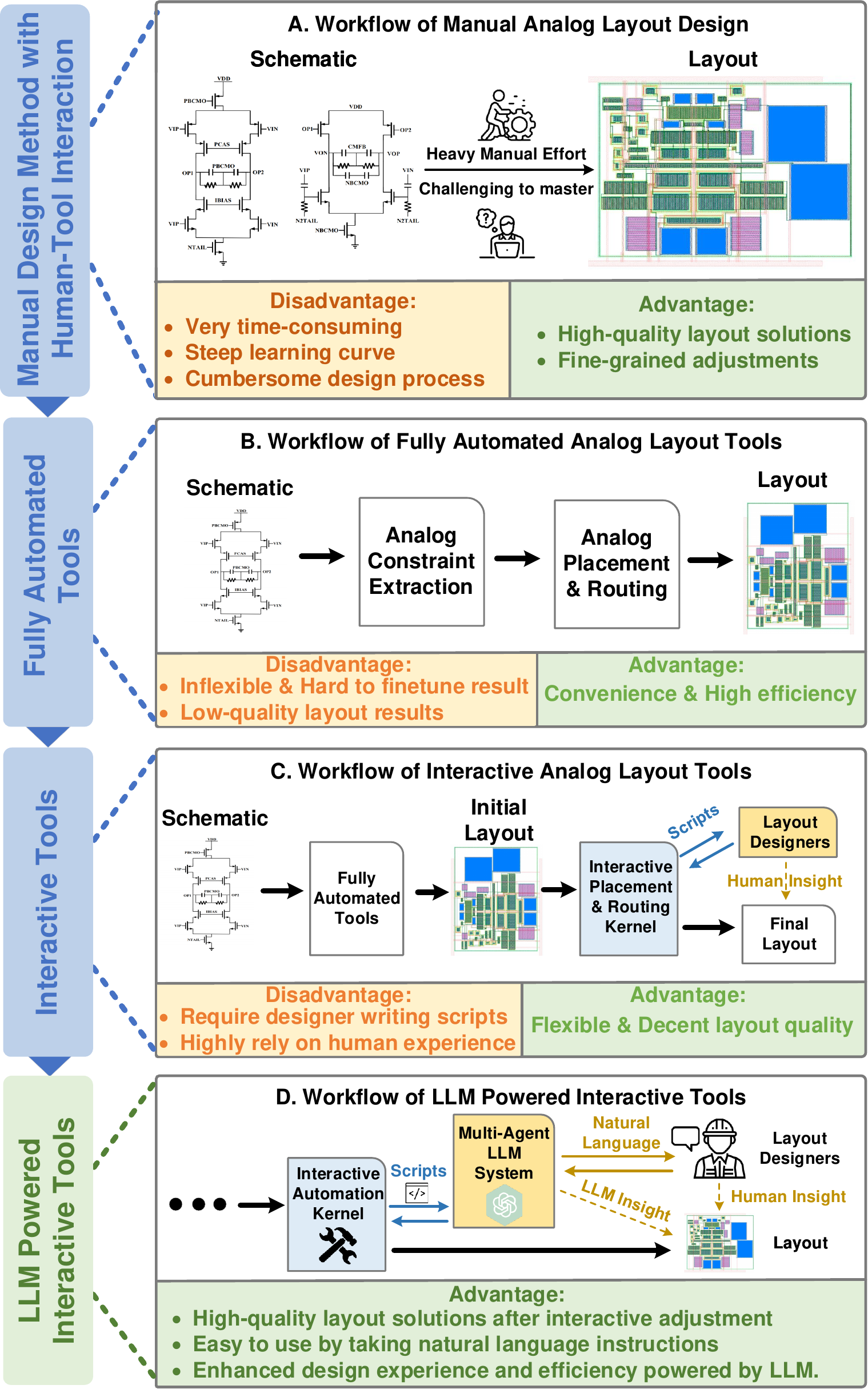}
    \caption{Comparison of workflows in analog layout automation, highlighting key advantages and disadvantages of manual, fully automated, interactive, and LLM-powered interactive tools (LayoutCopilot).}
    \label{fig:Intro}
    \vspace{-0.7cm}
\end{figure}

\begin{sloppypar}
    Analog layout design is a critical phase in analog circuit design that relies heavily on the manual effort of skilled designers\chg{, as shown in Figure~\ref{fig:Intro} A}. This dependence is largely due to the complexity of analog circuit performance models and the additional considerations such as symmetry, matching, signal flow, and other constraints, which pose significant challenges in generating high-quality layouts with superior performance. Therefore, efficiently generating high-quality layouts of analog circuits is a major challenge for both commercial tools and academic research.
\end{sloppypar}


\begin{sloppypar}
    Commercial analog layout design platforms like Cadence Virtuoso~\cite{virtuoso2023} mainly provide manual layout drawing interfaces and a few interactive layout automation functionalities. However, the learning curve for such platforms is very steep due to complicated buttons, shortcut keys, and scripting commands for manipulation. Their interactive layout automation functionalities require designers to type all the constraints, which is extremely tedious and rarely adopted by designers in practice. 
\end{sloppypar}

\begin{sloppypar}
    Academic studies have focused on both fully automated analog design methodologies and interactive automation. Over the past four decades, a series of analog placement \& routing algorithms have been proposed to automate layout design. These algorithms explore various methods and perspectives, including traditional algorithmic approaches~\cite{391116, 510537, malavasi1990routing,xiao2010practical,xu2017hierarchical}, domain-knowledge based methods~\cite{basaran1993latchup, ou2013simultaneous, martins2016current,xu2019device, ho2013coupling}, and the integration of machine learning techniques~\cite{li2020exploring, li2020customized, gusmao2020semi}. All are aimed at boosting the efficiency and performance of final layout results, paving the way for the development of fully automated tools. Recent advancements in fully automated tools for analog layout design such as ALIGN~\cite{kunal_align_2019, dhar_align_2021, sapatnekar_align_2023} and MAGICAL~\cite{xu_magical_2019, chenMAGICALOpenSource2021, chen_magical_2021}, have significantly improved design efficiency in analog layout generation. However, despite their advancements, these tools often fall short of accommodating the highly customized needs of analog layout design\chg{, as shown in Figure~\ref{fig:Intro} B}. To address this, interactive analog layout editing tools~\cite{gaoInteracplacement,gao_interactive_2023} have been developed to allow designers to modify and optimize layouts more easily.  While these tools offer enhanced flexibility in layout design, similar to commercial tools, they also introduce a new usability challenge as designers must master complex command sets and effectively apply them in circuit optimization\chg{, as shown in Figure~\ref{fig:Intro} C}. 
\end{sloppypar}

\begin{sloppypar}
    The above usability challenges come from the fundamental gap between human natural language and machine language. It is not easy to convert both designers’ concrete tasks and abstract design intents into executable commands for machines. Recent advances in large language models (LLMs) bring a new opportunity to bridge the gap. Recently prevailing models like GPT-3~\cite{brown2020language}, Llama~\cite{touvron2023llama, touvron2023llama2}, GPT-4~\cite{openai_gpt-4_2024}, and Claude~\cite{claude2, 2005electromigrationaware} have demonstrated remarkable capabilities in not only understanding and generating human-like text but also reasoning and comprehending abstract domain knowledge, paving the way for innovative applications across various domains. This evolution, along with advances in knowledge retrieval techniques~\cite{lewis_retrieval-augmented_nodate} and the proven effectiveness of multi-agent approaches in tackling complex reasoning tasks~\cite{wu_autogen_2023, rasal_llm_2024, hong_metagpt_2023}, sets a new stage for enhancing human-tool interaction in intricate technical tasks. Such advancements have led to diverse applications within EDA~\cite{liu_rtlcoder_2024, thakur_verigen_2024, lu_rtllm_2023, liu_verieval_2023, tsai_rtlfixer_2024, liu_chipnemo_2024, he_chateda_2024, wang2024chatpattern}, but mainly focused on digital circuits, where mature automation solutions have been widely adopted without heavy human-tool interaction. Thus, the field of analog layout design presents a unique opportunity as it has not yet fully explored using LLMs to tackle layout tool usability challenges.
\end{sloppypar}

\begin{sloppypar}
    In this paper, we introduce \textsl{LayoutCopilot}, a pioneering multi-agent collaborative framework powered by LLMs for interactive analog layout design. LayoutCopilot is proficient in processing concrete layout adjustment requests, translating them into executable commands that reduce the learning curve associated with interactive tools\chg{, as shown in Figure~\ref{fig:Intro} D}. Additionally, it leverages a comprehensive knowledge base to generate practical suggestions according to the designer's high-level design intents, thus addressing the usability challenges posed by the inherent complexity of analog design. The main contributions of this paper can be summarized as follows: 
\end{sloppypar}

\begin{itemize}
    \item  To the best of our knowledge, LayoutCopilot is the first LLM-powered interactive analog layout design framework, offering a novel approach to enhance design flexibility and efficiency in analog circuit design.
    \item We bridge the interaction gap between designers and analog layout tools with a framework powered by LLMs, overcoming the usability challenge of layout tools and refining the methodology of interactive layout design.
    \item We utilize a multi-agent collaborative framework that progressively transforms designer requirements into executable commands through coordinated efforts among multiple LLM agents, achieving high accuracy.
    \item Bulk testing and experiments on real-world analog designs have shown that LayoutCopilot can accurately address the designer's concrete layout adjustment requests and offer actionable suggestions to complete layout optimization based on their high-level design intents.
\end{itemize}

\begin{sloppypar}
The rest of the paper is organized as follows. 
Section~\ref{sec:Preliminary} describes the background; 
Section~\ref{sec:Algorithm} explains the detailed implementation; 
Section~\ref{sec:Results} demonstrates the results; 
Section~\ref{sec:Conclusion} concludes the paper. 
\end{sloppypar}


%% file: doc/prelim.tex
\section{Preliminaries}
\label{sec:Preliminary}

This section reviews the background concepts of our study, including the integration of LLMs with EDA, prompt engineering, multi-agent collaboration, and the interactive placement and routing in analog layout design,  additionally outlining the scope of \textsl{LayoutCopilot}.

\subsection{Integrating LLMs into EDA Applications}

\begin{sloppypar}
    Recent advancements in pre-trained large language models (LLMs) have unveiled new opportunities for enhancing EDA applications. The ability of LLMs to generate human-like text and understand complex concepts makes them ideally suited for integration into EDA tasks, ranging from auto-generating Hardware Description Language (HDL) code to facilitating interactive design workflows through conversational interfaces.
\end{sloppypar}

\begin{sloppypar}
    One segment of research has focused on harnessing LLMs to tackle textual or language-based tasks in EDA autonomously. This includes efforts to auto-generate HDL code using tools like RTLCoder and VeriGen~\cite{liu_rtlcoder_2024, thakur_verigen_2024}, along with benchmarks like RTLLM and VerilogEval for assessing these capabilities~\cite{lu_rtllm_2023, liu_verieval_2023}. Additionally, RTLFixer~\cite{tsai_rtlfixer_2024} is exploring automated debugging and code repair, whereas ChipNeMo~\cite{liu_chipnemo_2024} serves as an engineering assistant chatbot, facilitating EDA script generation and bug analysis. 
\end{sloppypar}

\begin{sloppypar}
    Another pathway seeks to help traditional design flows and existing EDA tools with LLMs, thus enhancing their accessibility and ease of use. This is exemplified by ChatEDA~\cite{he_chateda_2024} and ChatPattern~\cite{wang2024chatpattern}, which typically feature a conversational interface that allows designers to express their needs in natural language, thereby facilitating the indirect manipulation of EDA tools via LLMs. Introducing conversational interfaces powered by LLMs helps tackle long-standing usability challenges in EDA tools. Inspired by conversational interfaces in EDA tools, we applied LLMs to interactive analog layout design. Considering the custom nature of analog circuits, LayoutCopilot not only facilitates direct command execution but also generates actionable suggestions based on high-level design intents, streamlining the design process.
\end{sloppypar}




\subsection{Prompt Engineering}

\subsubsection{Introduction to Prompt Engineering}

\begin{sloppypar}
    \textsl{Prompts} are natural language instructions that provide context to guide the generative language model.  \textsl{Prompt engineering} is the process of leveraging prompts to enhance model efficacy without modifying parameters of the core model~\cite{sahoo2024systematic}. 
    The fundamental concept behind prompt engineering involves formulating queries or instructions in a manner that allows the model to understand and respond accurately, leveraging its pre-trained knowledge. Additionally, prompts can be constructed or automatically generated to convey information beyond the pre-trained \chg{datasets} to the model, thus flexibly adjusting the model's responses and enabling the integration of external knowledge bases. As language models based on the Transformer architecture become increasingly prevalent across various applications, crafting effective prompts that guide these models to generate useful and accurate outputs has become critically important.
\end{sloppypar}
\begin{sloppypar}
    Compared to other methods of improving LLM performance, \chg{prompt engineering} offers a lightweight solution for utilizing large models to solve real-world problems. Unlike traditional methods that require extensive data set creation, fine-tuning, and repetitive adjusting to enhance a model's performance on \chg{specific tasks. Prompt engineering} allows practitioners to directly transmit knowledge and methodologies to the model through carefully designed prompts. This approach is especially useful in fields where data acquisition and cleansing are challenging, such as EDA.
\end{sloppypar}

\subsubsection{Application and Advanced Techniques}

\begin{sloppypar}
    \chg{In-context learning is a paradigm enabling language models to perform tasks by learning from only a few examples provided as demonstrations~\cite{dong2022survey}. One approach of prompt engineering is to include these examples directly in the prompt, which can be categorized based on the number of examples provided: zero-shot, one-shot, and few-shot learning~\cite{LanguageModelsareFewShotLearners}. While prompt engineering often involves providing such examples, it is not limited to this method. Another approach leverages abstract language to guide LLMs' behavior in generating responses, using techniques like chain-of-thought and self-refinement to further enhance model capabilities.}
\end{sloppypar}

\begin{sloppypar}
    Researchers have developed advanced prompt engineering techniques to boost LLMs' problem-solving abilities. Building on the `few-shot learning' approach\cite{LanguageModelsareFewShotLearners}, researchers have discovered that an `abstract description' of examples can sometimes be sufficient, eliminating the need for direct conversational examples. Here are a few examples of the above:
    \begin{itemize}
        \item \textsl{Chain-of-thought} guides LLMs to solve problems through a series of intermediate steps, which improves LLMs' reasoning ability by inducing the model to articulate a multi-step problem-solving process that mimics human reasoning~\cite{ChainOfThoughtPromptingElicitsReasoninginLargeLanguageModels}.
        \item \textsl{Least-to-most prompting} Least-to-most prompting simplifies complex problems into manageable subtasks, tackled sequentially, significantly boosting the model's ability to solve intricate reasoning tasks~\cite{LeastToMostPromptingEnablesComplexReasoningInLargeLanguageModels}.
        \item The \textsl{Self-refine strategy} enhances \chg{LLMs'} performance through an iterative process where the model critiques and revises its solutions, thereby refining the quality of its responses \cite{SelfRefineIterativeRefinementwithSelfFeedback}. 
    \end{itemize}
    To handle complex reasoning tasks in our application, LayoutCopilot integrates various Prompt Engineering techniques, including those mentioned above. Each agent is tailored according to the specific task it performs to achieve better performance. Details about the structure of prompts are presented in Section~\ref{sec:Algorithm}. 
\end{sloppypar}

\subsubsection{Automatic Prompt Generation and Retrieval-Augmented Generation (RAG)}

\begin{figure}[tb]
    \centering
    \includegraphics[width=0.45\textwidth]{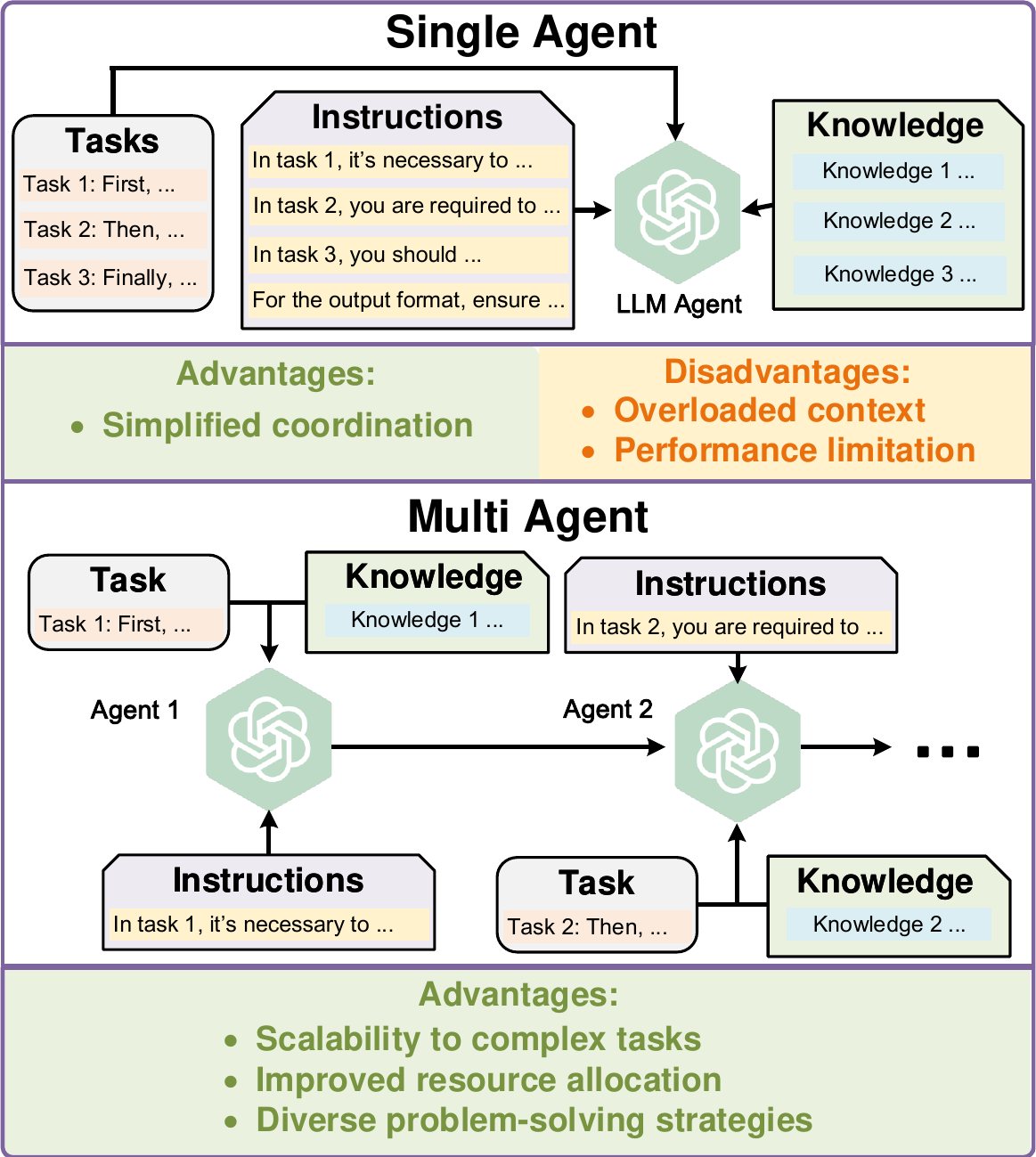}
    \caption{System comparison: overcoming single agent limitations through multi-agent collaboration.}
    \label{MultiAgent}
    \vspace{-0.5cm}
\end{figure}

\begin{sloppypar}
    Apart from manually configured prompts, there are also techniques for automatic prompt generation. A classic example is Retrieval-Augmented Generation (RAG)~\cite{RetrievalAugmentedGenerationForKnowledgeIntensiveNLPTasks}, which enhances the model's interaction with external knowledge bases. As previously mentioned, prompts often contain a few examples, thus aligning with the `few-shot' approach. RAG allows these examples to be automatically retrieved from a database or knowledgebase, providing an interface through which LLMs can access external knowledge and align their responses accordingly. However, the retrieved context, combined with system instructions and designer requests, is fed as text input to the \chg{LLMs}, which could lead to prompt dilution if the context is overly extensive. This can result in a decrease in response quality, an issue we have observed in our experiments. To address this, we employ a multi-agent methodology to segment different tasks among several LLM agents, ensuring that knowledge retrieval does not compromise the performance of other system components. By separating agents with distinct responsibilities, we maintain system efficiency even as we integrate context-extensive retrieval technologies into our framework.
\end{sloppypar}



\subsection{Multi-Agent Collaboration with LLMs}

\begin{sloppypar}
    The objective of multi-agent collaboration is to enable multiple autonomous agents to effectively collaborate towards a shared goal~\cite{zhang2024proagent}. Figure~\ref{MultiAgent} illustrates the contrast between the capabilities of single and multi-agent systems, highlighting how multi-agent collaboration significantly enhances the system's capacity by leveraging the specialized expertise of each agent and preventing task interference. This allows the system to manage a considerably larger workload than possible without such specialization, leading to improved efficiency and output quality~\cite{xi2023rise}.
\end{sloppypar}

\begin{sloppypar}
    After role-playing capabilities were introduced into communicative agent interactions by~\cite{li2023camel}, \cite{talebirad2023multi} proposed a comprehensive LLM-based multi-agent collaboration framework that demonstrated efficiency enhancements in handling complex tasks. Based on these works, multi-agent methodologies have been successfully deployed in various applications, proving their effectiveness in scenarios such as text understanding, reasoning, mathematics, coding, and tool utilization~\cite{rasal2024llm, wu2023autogen, chen2023agentverse}. Furthermore, several improvements have been proposed to enhance multi-agent collaboration, including~\cite{hong2023metagpt}, which transfers professional knowledge and management experience to LLM agents for more structured collaboration, and~\cite{zhang2024proagent}, which improves cooperation by enabling agents to predict their collaborators' actions.
\end{sloppypar}

\begin{sloppypar}
    In LayoutCopilot, we have developed a multi-agent collaborative framework that divides complex tasks into specialized subtasks.
    This strategic division of labor ensures that each agent can operate within its expertise without diluting the prompt or compromising other tasks' performance.
\end{sloppypar}


\subsection{Interactive Analog Layout Design}

\begin{sloppypar}
    Analog layout automation has engaged many researchers in recent years and several basic methodologies have emerged~\cite{xiao2010practical,xu2017hierarchical,ou2013simultaneous, martins2016current,xu2019device, ho2013coupling,li2020customized,gusmao2020semi,kunal_align_2019, dhar_align_2021, sapatnekar_align_2023,xu_magical_2019, chenMAGICALOpenSource2021, chen_magical_2021,gaoInteracplacement,gao_interactive_2023}.  The analog placement problem can be formulated into a nonlinear optimization problem. The most common objective function is half-perimeter wire length (HPWL), indicating the performance of wire length. Different from digital placement, analog placement will consider more constraints such as symmetry, array, etc. The analog routing problem is a pathfinding problem that can be solved by a typical shortest path algorithm (A-star). 
\end{sloppypar}

\begin{sloppypar}
    An interactive analog layout design framework offers flexible adjustments to the layouts while relieving designers from tedious manual layout drawing. By introducing high-level interactive controllers, the framework can leverage placement and routing kernels to automatically adjust the layout. Based on the interactive layout design framework, designers can add arbitrary placement constraints and refine the routing solution according to their design experience. After several adjustments, designers can finally obtain a high-quality layout. 
\end{sloppypar}

\subsection{The Scope of LayoutCopilot}

\begin{figure}[tb]
    \centering
    \includegraphics[width=0.3\textwidth]{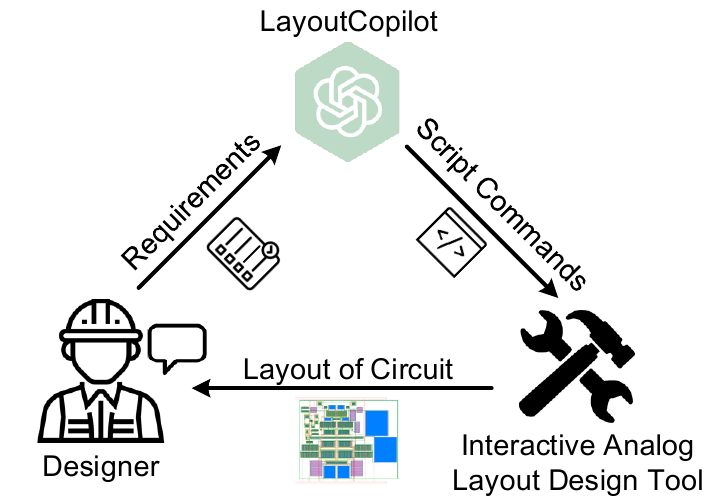}
    \caption{Illustration of LayoutCopilot's functionalities.}
    \label{Scope}
    \vspace{-0.5cm}
\end{figure}

\begin{sloppypar}
    LayoutCopilot functions as a multi-agent collaborative framework designed to enhance the interaction between designer and layout tools, illustrated by Figure~\ref{Scope}. Its primary objective is to interpret design intents expressed in natural language and generate actionable commands that effectively interact with layout tools.
\end{sloppypar}

\begin{sloppypar}
    \textbf{Problem 1.} \textsl{(Objective of LayoutCopilot)} Develop an LLM-powered framework that accepts both high-level design intents and concrete requests in natural language from designers and then transforms them into precise executable commands. These commands aim to manipulate the layout tools directly, ensuring that the design intents are accurately translated into layout adjustments.
\end{sloppypar}

\begin{sloppypar}
    \chg{The main challenges in developing LayoutCopilot involve coordinating multiple tasks within the system, such as high-level design intent parsing, device-level layout adjustments, and generating accurate commands for layout tools—each of which differs in complexity and focus. These tasks range from design intent parsing to precise command generation, which can be challenging for a single-agent system to manage effectively. Additionally, integrating a knowledge base can consume substantial model attention and interfere with unrelated tasks. To address these issues, LayoutCopilot employs a multi-agent system that dedicates specific agents to fixed tasks, enhancing focus and response quality. This setup also allows agents to draw from the knowledge base while minimizing adverse effects on other tasks, thereby enhancing the system’s overall accuracy and efficiency in translating design intents into actionable layout commands.}
\end{sloppypar}

%% file: doc/algo.tex
\begin{figure*}[tb]
    \centering
    \includegraphics[width=0.96\textwidth]{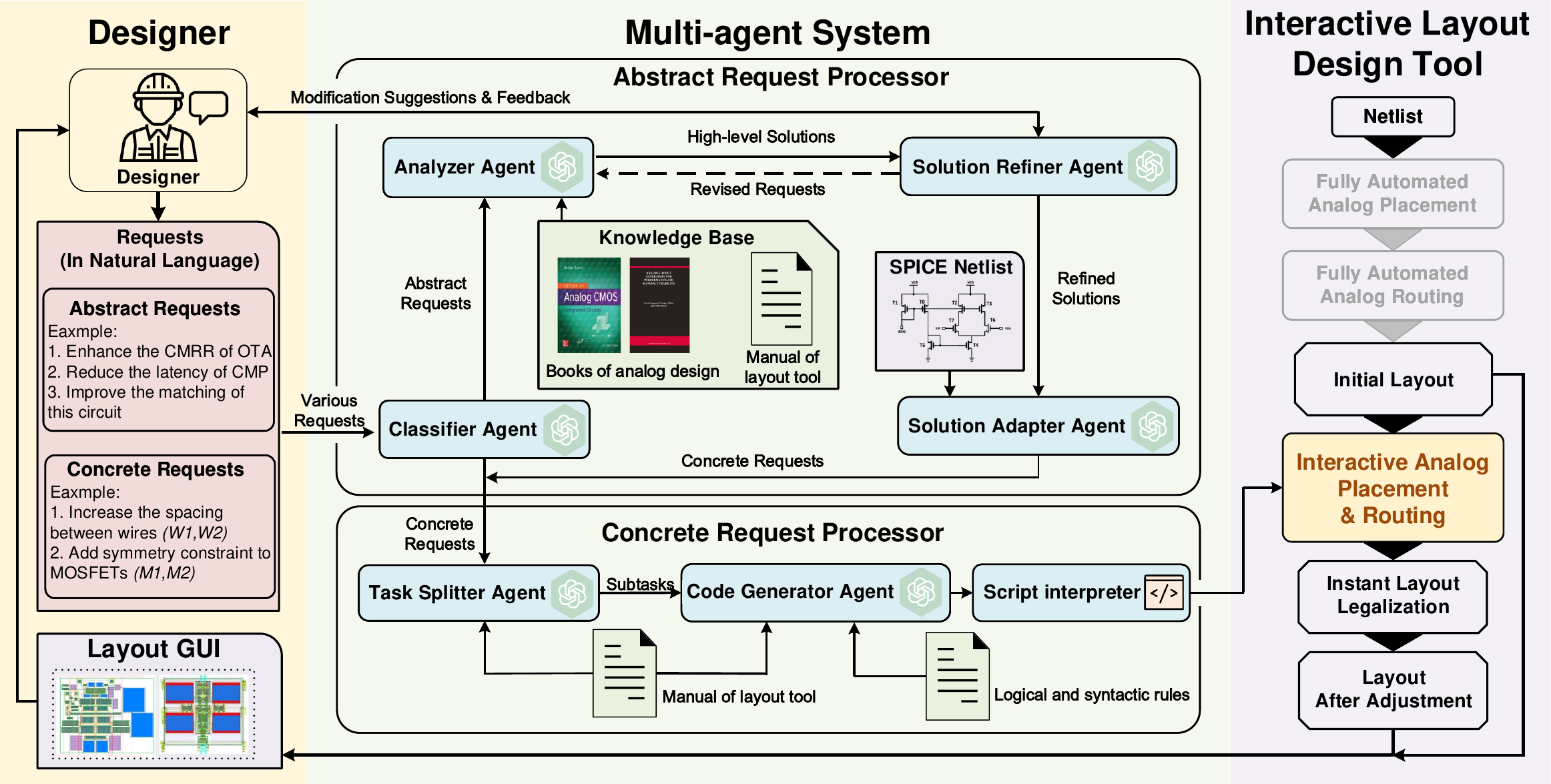}
    \caption{Overview of LayoutCopilot: a multi-agent framework that interprets natural language design intents through abstract and concrete request processors, coordinating agents to execute precise layout adjustments for interactive placement and routing.}
    \vspace{-0.5cm}
    \label{FlowFramework}
\end{figure*}

\section{LayoutCopilot Framework}
\label{sec:Algorithm}

\begin{sloppypar}
    This section introduces the architecture and functionality of LayoutCopilot, a multi-agent collaboration framework enhanced by LLMs for interactive analog layout design. LayoutCopilot enhances the interaction between designers and layout tools by understanding complex requirements in natural language, utilizing a knowledge base for solution generation, and automating layout design. It consists of two primary components: the Abstract Request Processor and the Concrete Request Processor, as illustrated in Figure~\ref{FlowFramework}. The Abstract Request Processor initiates the processing of requests, transforming them into concrete requests that adhere to the layout tool manual. These concrete requests are then transferred to the Concrete Request Processor, which generates executable commands to facilitate layout editing. This workflow simplifies interactions between layout tools and designers, enhancing the overall design process. Detailed discussions on each component's functionality and the rationale for their design are provided in the subsequent subsections.
\end{sloppypar}

\subsection{Abstract Request Processor}

\begin{sloppypar}
    The Abstract Request Processor plays a key role in LayoutCopilot, managing and processing diverse requests from designers. It performs a series of functions including classification of requests, retrieval of relevant knowledge, iterative adjustments in collaboration with designers, and analyzing netlists to transform abstract requests into concrete ones that align with layout tool manuals. This component is powered by four specialized LLM agents, each dedicated to a specific aspect of the processing pipeline: classification, analysis, solution refinement, and solution adaptation, as shown in Figure~\ref{FlowFramework}. 
\end{sloppypar}

\begin{sloppypar}
    Before we discuss the construction of specific agents, let us first explain why we use multiple LLM agents to build such a complex system instead of relying on a single LLM agent to solve all problems. The reason we divide tasks among multiple agents instead of relying on a single agent is to mitigate the "dilution" of prompts, and the distraction of the \chg{LLMs'} attention. In our experiments, we noted that each LLM agent can effectively handle only a limited context. When provided with long and complex instructions as prompts, the performance of LLM handling a task decreases compared to when given shorter instructions. Furthermore, if the instruction includes multiple tasks, they tend to "dilute" each other, resulting in poorer performance compared to when the instruction contains only a single task. This phenomenon is similar to the \chg{LLMs'} limited attention being split among the tasks. This issue is particularly pronounced when involving long-text knowledge retrieval. When transmitting knowledge bases to the \chg{LLMs}, they are input either directly as prompts, or through the LLM provider's retrieval interface, which also compresses the content before entering the prompt. In both cases, the prompt's content increases. Even after compression, the knowledge base's content remains large, meaning that if an agent accesses the knowledge base, it can only effectively handle tasks around this knowledge base. Other tasks' performance would suffer due to prompt "dilution."
\end{sloppypar}

\begin{sloppypar}
    Therefore, we adopt a multi-agent structure, dividing tasks into several independent parts and assigning them to different agents to ensure each task is handled efficiently. This approach helps maintain the effectiveness of each agent by preventing the dilution of prompts and allows the system to manage complex and varied requests more effectively. Details of these agents will be shown as follows:
\end{sloppypar}

\subsubsection{Classifier Agent}
\begin{sloppypar}
    \scdchg{At the head of the analyzer lies a task classifier agent, working as a filter to determine designers' requirements as either concrete or abstract requests based on the manual of layout tools.} Concrete tasks are identified by their direct translatability into commands or combinations explicitly supported by the layout tools, for example, `add symmetry between M6 and M7'. Conversely, abstract requests are recognized for their high-level, conceptual nature, necessitating a comprehensive analysis grounded in the netlist of circuits and knowledge of layout design to be deconstructed into executable steps, such as `Enhance the matching' and `Improve the CMRR'. Building on these criteria, an LLM agent executes the task classification. Once classified, the tasks are directed into two distinct flows: concrete tasks are sent directly to the Concrete Request Processor for immediate processing. At the same time, abstract requests undergo a series of steps to be transformed into concrete requests before they are processed in the same manner as concrete requests. \chg{\scdchg{Without it, abstract requests may be directly sent to the Concrete Request Processor, leading to incorrect task processing.}}
\end{sloppypar}

\begin{figure*}[tb]
    \centering
    \includegraphics[width=0.96\textwidth]{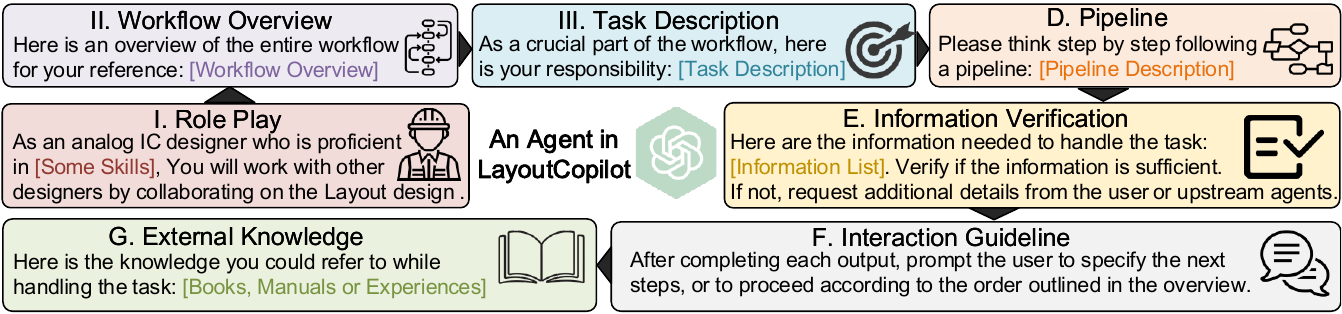}
    \caption{Illustration of the configuration for a single agent in LayoutCopilot.}
    \vspace{-0.5cm}
    \label{SingalAgent}
\end{figure*}

\subsubsection{Analyzer Agent}
\begin{sloppypar}
    The Analyzer Agent is dedicated solely to the analysis and knowledge retrieval task, which is crucial given LLMs' limitations with long context lengths. Tasks that require extensive context can use up much of this capacity, potentially diluting prompts and reducing the effectiveness of other tasks as explained above. To prevent these issues, this agent focuses exclusively on knowledge retrieval. The Analyzer Agent operates within a vast knowledge base that spans specialized literature on analog circuits and layout design, layout tool manuals, and archives of previous tasks and solutions. When transmitting the knowledge bases to the \chg{LLMs}, they are input either directly as part of prompts or through the LLM provider's retrieval interface.
\end{sloppypar}

\begin{sloppypar}
    The Analyzer Agent crafts high-level solutions based on the retrieved knowledge, functioning similarly to a skilled architect drafting a blueprint before construction and outlining strategic approaches to the task. For example, when tasked with 'enhancing the CMRR of an OTA,' the agent might suggest optimizing component placement to enhance symmetry and rerouting connections to improve parasitic matching, all tailored to the specific capabilities of the chosen layout tools. The high-level solutions are then sent to the Solution Refiner Agent for refinement and adjustment based on the designer's modification suggestions. If the designer is dissatisfied with the results, the Analyzer Agent will be re-invoked. The revised request, which includes the designer's modification suggestions, will be sent back from the Solution Refiner Agent as the new input for the Analyzer Agent to generate optimized high-level solutions. This mechanism ensures that the system can incorporate designer feedback to improve design solutions for more accurate and targeted results.
\end{sloppypar}

\subsubsection{Solution Refiner Agent}
\begin{sloppypar}
    Considering analog designers' rich experience and highly customized requirements, enhancing their involvement in the workflow is crucial. The Solution Refiner Agent facilitates this by presenting high-level solutions to designers along with detailed evaluations and explanations. This allows designers to understand the rationale behind each solution, assess its feasibility, and make informed decisions. The designer can select from these solutions, modify them as needed, or provide additional feedback to refine the results further. If designers find the solutions unsatisfactory, they have the option to add more details or revert to the Analyzer Agent for adjustments. This iterative feedback process ensures that high-level solutions are continuously refined based on the designer's expertise, ultimately aligning the final layout with the designer's specific requirements and project needs. Once refined to the designer's satisfaction, the solutions are then passed to the Solution Adapter Agent. \chg{To illustrate this process more clearly, the following example dialogue demonstrates the interactions between the Solution Refiner Agent, other agents, and the designer. This example also shows the interaction process among agents; dialogues for other agents follow a similar structure and are thus omitted for brevity. }
\end{sloppypar}


\begin{tcolorbox}[colback=gray!5!white, colframe=gray!75!black, title=Example Dialogue: Solution Refiner Agent Interactions, fonttitle=\bfseries, boxrule=0.75pt, arc=2mm, left=3pt, right=3pt, top=3pt, bottom=3pt]
    \scriptsize
    \textbf{To Designer:}
    \begin{quote}
    Improving the performance of your OTA layout can involve...\\
    Here are some potential high-level solutions:
    \begin{enumerate}
    \item Enhance Symmetry with \texttt{symAdd} ...
    \item Improve Matching with \texttt{deviceMove} and \texttt{deviceSwap} ...
    \item Reduce Parasitics with \texttt{wireWidth} and \texttt{wireSpacing} ...
    \item Prevent Crosstalk with \texttt{wireSpacing} ...
    \item Improve Routing with \texttt{netPriority} and \texttt{netTopology}...
    \end{enumerate}
    Please select some of the high-level solutions you would like to execute, or provide any modifications or suggestions. Once confirmed, we will convert these into concrete requests for implementation.
    \end{quote}
    
    \textsl{\textbf{Designer:} I prefer to enhance the symmetry.}
    
    \textbf{To Designer:}
    \begin{quote}
    Great choice! You have selected \textbf{Enhance Symmetry} with the \texttt{symAdd} function.
    
    I will now forward this task to the Solution Adapter for further interaction...
    \end{quote}
    
    \textbf{To Adapter:}
    \begin{quote}
    We have received a selection from the Designer to \textbf{Enhance Symmetry} using the \texttt{symAdd} function. The following are detailed requirements for enhancing layout symmetry:
    \begin{enumerate}
    \item \textbf{Confirm the Details:} Gather any necessary parameters or specific configurations needed for the \texttt{symAdd} function from the netlist and the Designer.
    
    \item \textbf{Engage with the Designer:} If any clarifications or additional inputs are required, reach out to the Designer for feedback or modifications.
    
    \item \textbf{Prepare Concrete Requests:} Once all details are confirmed, convert the task into actionable concrete requests with sufficient and correct parameters for further processing.
    \end{enumerate}
    \end{quote}
\end{tcolorbox}

\subsubsection{Solution Adapter Agent}
\begin{sloppypar}
    \scdchg{The Solution Adapter Agent is the final component of the Abstract Request Processor, tasked with converting the requests refined in the previous steps into concrete, actionable requests.} Depending on the complexity and requirements, netlists are either communicated directly to the \chg{LLMs} during dialogues or stored in a knowledge base accessed via the \chg{LLMs'} retrieval interface. This agent meticulously analyzes the netlist to extract specific parameters following the directives from the refined solutions. For instance, when receiving a refined solution that specifies `enhance the symmetry of differential pairs and capacitors,' the Solution Adapter Agent locates these components in the netlist by examining their connectivity and device parameters. It then formulates concrete requests to apply symmetry constraints to the identified components, such as `add symmetry between M6 and M7.' \scdchg{Refined requests may lack the specificity necessary for implementation without the Solution Adapter Agent, resulting in incomplete or incorrect actions.}
\end{sloppypar}

\begin{sloppypar}
    Apart from extracting parameters from the netlists, the Solution Adapter Agent also ensures that the generated requests are compatible with the layout tools, adhering to the specific syntax and operational requirements of these tools. This compatibility is crucial for translating refined solutions into actionable steps, bridging the gap between design intent and practical implementation. \chg{Without this agent, requests sent to the Concrete Request Processor may fail to be processed, resulting in errors or incomplete execution of generated commands.}
\end{sloppypar}

\subsection{Concrete Request Processor}

\begin{sloppypar}
    The Concrete Request Processor is dedicated to accurately translating concrete tasks into executable commands that fulfill designer requirements via layout tools. The process begins by decomposing concrete tasks into subtasks, each corresponding to a command detailed in the manual of layout tools, akin to the processes used in~\cite{he_chateda_2024} and~\cite{wang2024chatpattern}. However, merely adopting these methods does not guarantee that the generated commands will execute correctly or achieve the desired results in this scenario. Distinct from previous works, our approach incorporates role-playing and task decomposition techniques while conveying the syntactic and logical rules dictated by the layout tools' manuals to the Code Generator Agent. This enrichment ensures that the command sequences generated by the agent maintain high accuracy in syntax and logic, especially when dealing with complex input requests. This strategic enhancement is committed to improving the precision and reliability of the process, closely aligning with the specific operational requirements of the layout tools.
\end{sloppypar}

\vspace{-0.35cm}
\subsection{Agent Configuration and Prompt Design}\label{subsec: agent configuration}

\vspace{-0.0cm}
\begin{sloppypar}
    Each agent in the LayoutCopilot framework is equipped with prompts tailored to its designated task. These prompts are crafted using the prompt engineering techniques described in Section \ref{sec:Preliminary}, chosen to enhance the overall expressive capabilities of the agents based on experimental results, as illustrated in Figure \ref{SingalAgent}. 
\end{sloppypar}

\begin{sloppypar}
    The seven prompt sections are described in detail as follows: 
    \begin{itemize}
        \item \textsl{I. Role Play} encourages the agent to adopt a persona that helps simulate a realistic and professional scenario for task handling.
        \item \textsl{II. Workflow Overview} provides agents with an understanding of their roles within the whole system, highlighting how their outputs influence subsequent stages of the design process.
        \item \textsl{III. Task Description} delineates the specific responsibilities and objectives of the agent, ensuring focused and relevant task execution.
        \item \textsl{IV. Pipeline} guides agents through a chain-of-thought approach to tackle the assigned tasks efficiently. This enhances the agents' reasoning capabilities and directs them to break down complex tasks into manageable sub-tasks, improving their ability to handle intricate problems.
        \item \textsl{V. Information Verification} instructs the agent to verify the completeness of the input before proceeding, prompting for additional information if required. This acts as a self-refine strategy, effectively preventing errors arising from incomplete or inaccurate inputs.
        \item \textsl{VI. Interaction Guideline} directs the agent on interacting with the designer and other agents, ensuring that the entire process remains cohesive and efficient.

        \item \textsl{VII. External Knowledge} \scdchg{enables each agent to access resources tailored to its role, enhancing the output with domain-specific insights. The knowledge provided to each agent is specifically aligned with their functional requirements. Books on analog and layout design~\cite{razavi2001design, lampaert1999analog} are provided exclusively to the Analyzer Agent to build a comprehensive knowledge base without overwhelming other agents, as these extensive materials could interfere with their tasks. Specifically, these books are provided in text form, serving primarily as a general roadmap.  Additionally, all agents have access to layout tool manuals, command lists (as shown in Table I), and examples of request processing. These resources are provided in a structured text format with annotations to help the LLM understand and use the information effectively. This ensures compatibility with the tools' syntax and guides precise and consistent task execution across the framework.}


    \end{itemize}
          
\end{sloppypar}

\begin{sloppypar}
    \chg{Regarding the detail of the prompt, we present the Pipeline section of the Solution Refiner Agent's prompt as an example, given that it is the most complex agent.}
    
    \begin{tcolorbox}[colback=gray!5!white, colframe=gray!75!black, title=Solution Refiner Agent Prompt: Pipeline Section, fonttitle=\bfseries, boxrule=0.75pt, arc=2mm, left=3pt, right=3pt, top=3pt, bottom=3pt]
    \scriptsize
    Please proceed according to the following pipeline, depending on the source of the request: 
    
    *If the request comes from the Analyzer Agent, follow Pipeline A; if the request comes from the Designer, follow Pipeline B.* \\
    
    \#\#\# Pipeline A: Request from **Analyzer Agent**
    
    **Input:** A series of layout modification suggestions (i.e. high-level solutions) generated by the Analyzer Agent: {high\_level\_solutions} 
    
    **Steps:**
    
    - (a) Organize and Present High-Level Solutions: Compile the layout modification suggestions into a clear and structured format. Provide necessary explanations for each solution to help the Designer understand their implications and potential benefits.
    
    - (b) Engage with the Designer: Present the organized solutions to the Designer for feedback.
    **Output:** Please distinguish the output given to each recipient with "---To XXX---".
    
    - **To Designer:**  
      Organized high-level solutions with explanations for review.  \\
    
    \#\#\# Pipeline B: Request from **Designer**
    
    **Input:** Designer feedback on previously presented high-level solutions: {designer\_feedback}
    
    **Steps:**
    
    - (a) If the Designer is satisfied with a specific solution, compile the finalized solution for further interaction.
    
    - (b) If the Designer requests modifications, document the feedback in a structured format and send it back to the Analyzer Agent for refinement.
    
    **Output:** Please distinguish the output given to each recipient with "---To XXX---".
    
    - **To Adapter Agent (if Designer is satisfied):**    Refined solution(s) finalized by Designer.  
    
    - **To Analyzer Agent (if Designer requests modifications):**  
      Designer’s feedback and modification requests for further analysis.
    \end{tcolorbox}

    \chg{For requests from the Analyzer Agent, high-level solutions are organized and clearly presented to the Designer with explanations through Pipeline A, allowing for informed decision-making. When receiving feedback from the Designer, the prompt guides the agent to either send approved solutions to the Adapter Agent or to send modification requests back to the Analyzer Agent following Pipeline B. Each step includes explicit control flow, input and output descriptions as well as the chain of thought, supplying clarity and precision in the interaction across agents. }
\end{sloppypar}

\begin{sloppypar}
    In summary, these prompt components ensure that each agent operates not only as a specialist in its respective domain but also as a coherent part of the larger system, enhancing the overall functionality and efficiency of the LayoutCopilot. 
\end{sloppypar}

\subsection{Interactive Layout Editor}

\begin{figure}[tb]
    \centering
    \includegraphics[width=0.48\textwidth]{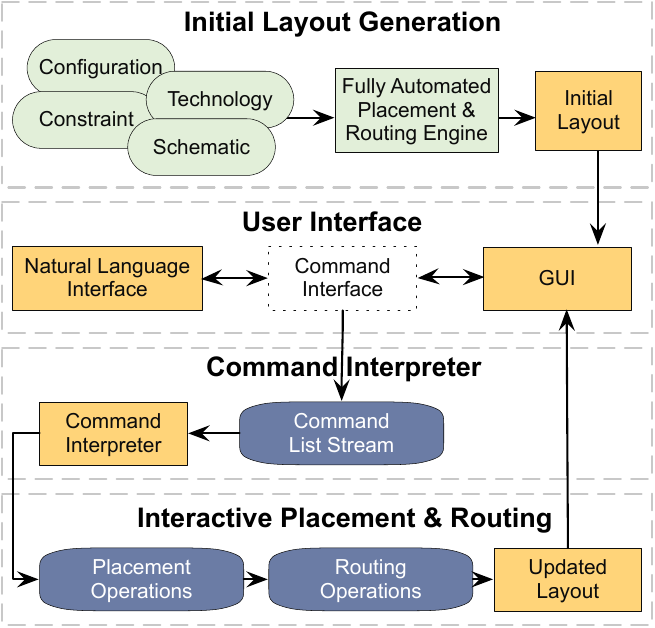}
    \caption{Overview of the interactive layout editor.}
    \label{InteractiveFlowFramework}
    \vspace{-0.3cm}
\end{figure}

\begin{table}[tb]
    \centering
    \caption{{Interactive command set.}}
    \resizebox{.48\textwidth}{!}{
\begin{tabular}{|c|c|}
\hline
\textbf{Command}   & \textbf{Parameters}  \\ \hline
\texttt{deviceMove}  & device $v_i$, destination location $(x, y)$ \\  \hline
\texttt{deviceSwap}  & devices $v_i, v_j$  \\  \hline
\texttt{arrayAdd} & a list of devices ${v_i}$, array shape ${x, y}$  \\  \hline
\texttt{arraySpace} & an array group $g_i$, horizontal space $x$, vertical space $y$ \\  \hline
\texttt{symAdd} & devices $v_i, v_j$, symmetry axis $A_k$  \\ \hline
\texttt{netRemove}  & net $n_i$  \\  \hline
\texttt{netReroute}  & net $n_i$  \\  \hline
 \texttt{wireWidth}  & wire $w_i$ of net $n_j$, new wire width $W$ \\  \hline
\multirow{2}{*}{\texttt{wireSpacing}} & wire $w_i$ of net $n_m$, wire $w_j$ of net $n_t$, spacing width $S$ \\                          
& wire $w_i$ of net $n_m$, device $v_j$, spacing width $S$   \\  \hline
\texttt{netPriority}  & a list of nets ${n_i}$ with their priorities  \\  \hline
\texttt{netTopology}  & net $n_i$, rough guide with points ${p_i}$  \\ \hline

\end{tabular}
}
    \vspace{-0.6cm}
    \label{tab:command_set}
\end{table}

\begin{sloppypar}
    Interactive layout editor bridges the gap between fully automated analog layout tools and real-world industrial applications. Our interactive analog editor is based on previous interactive works~\cite{gaoInteracplacement,gao_interactive_2023} with some extensions. Figure~\ref{InteractiveFlowFramework} shows the overview of the interactive layout editor. Before the interactive adjustment flow starts, an initial layout is given by fully automated tools. These tools take the netlists, technology files, and some basic constraints as the input. Based on the initial layout, users can start making interactive layout adjustments more intuitively through the user interface. The user interface consists of a natural language interface and a layout GUI. Compared to the previous script-based interface~\cite{gaoInteracplacement,gao_interactive_2023}, users can use natural language to adjust the analog layout directly. After converting the natural language into a list of commands in the command set shown in Table~\ref{tab:command_set}, the command interpreter will decompose each command into placement and routing operations. After the adjustment, the layout result will be shown in the GUI. Users can further adjust the layout according to the given result.
\end{sloppypar}

\begin{sloppypar}
    For placement adjustment, users are provided with seven different commands. Theoretically, arbitrary placement adjustments can be achieved by the combination of these commands. \texttt{move} command can move a device to a given location. \texttt{swap} command can swap the location of two given devices. For array-based adjustment, the \texttt{arrayAdd} command adds the array constraints to a group of devices and \texttt{arraySpace} adjusts the space between devices in the array group. Symmetry constraints can be added by \texttt{symAdd} command. With these commands, users can make placement adjustments easily. For instance, the Solution Refiner Agent decided earlier to enhance the symmetry of differential pairs, and the Solution Adapter Agent identified a differential pair \texttt{\{M6, M7\}}. Then the layout modification can be achieved by executing \texttt{symAdd M6 M7}, generated by the Concrete Request Processor. 
\end{sloppypar}

\begin{sloppypar}
    For routing adjustment, users are provided with six different commands. \texttt{remove} and \texttt{reroute} are basic commands for removing unsatisfied wires and rerouting them. \texttt{wireWidth} command can change the width of a given net or wire to improve the layout performance. \texttt{wireSpacing} command can adjust the spacing between two nets to avoid the signal cross-talk. \texttt{netPriority} command can designate a specific routing priority of nets. Manual guidance can be set by \texttt{netTopology} command, and the final routing solution will follow the guidance tightly. With these commands, users can accomplish arbitrary routing adjustments.   
\end{sloppypar}

%% file: doc/result.tex
\section{Experimental Results}
\label{sec:Results}



\begin{sloppypar}
    \chg{In this section, we will show experiments separately for concrete and abstract requests, showcasing LayoutCopilot's accuracy in handling basic tasks and its capacity to manage complex and comprehensive circuit optimization tasks.} LayoutCopilot is adaptable to various LLMs and layout tools, facilitating flexible deployment across different environments. In our experiments, we utilize a mature interactive analog layout design tool~\cite{gaoInteracplacement,gao_interactive_2023} for demonstration. Moreover, we employ different versions of prevailing LLMs to demonstrate the versatility of LayoutCopilot including GPT-3.5~\cite{brown2020language}, GPT-4~\cite{openai_gpt-4_2024}, and Claude-3~\cite{claude3}. 
    
\end{sloppypar}

\begin{table*}[tb]
\centering
\caption{{Sanity checks and comparison for single-agent with instruction vs. multi-agent with and without instruction.}}
\begin{tabular}{|c|c|c|c|c|c|c|c|c|c|c|c|c|}
\hline
\multirow{2}{*}{\textbf{Category}} & \multicolumn{4}{c|}{\textbf{Single-agent w/ Instruction}} & \multicolumn{4}{c|}{\textbf{Multi-agent w/o Instruction}} & \multicolumn{4}{c|}{\textbf{Multi-agent w/ Instruction}} \\ \cline{2-13} 

{} & {GPT-3.5} & {GPT-4} & {Claude-3} & \textbf{Avg.} & {GPT-3.5} & {GPT-4} & {Claude-3} & \textbf{Avg.} & {GPT-3.5} & {GPT-4} & {Claude-3} & \textbf{Avg.} \\ \hline
{Formatting}         & 71.14\% & 90.91\% & 99.25\% & 87.20\% & 82.00\% & 97.92\% & 99.83\% & 93.25\% & 95.38\% & 99.76\% & 99.92\% & \textbf{98.26}\% \\ \hline
{Validity}           & 91.36\% & 93.60\% & 95.44\% & 93.47\% & 96.88\% & 99.20\% & 98.40\% & 98.16\% & 98.24\% & 99.28\% & 98.88\% & \textbf{98.77}\% \\ \hline
{Syntax}             & 67.11\% & 88.87\% & 95.24\% & 83.74\% & 79.20\% & 95.28\% & 97.76\% & 90.75\% & 92.65\% & 97.20\% & 96.96\% & \textbf{95.60}\% \\ \hline
{Logic}              & 66.44\% & 83.04\% & 91.67\% & 80.38\% & 76.16\% & 93.44\% & 95.20\% & 88.27\% & 91.24\% & 94.24\% & 98.80\% & \textbf{94.76}\% \\ \hline
{\textbf{Overall}}     & 66.27\% & 82.91\% & 90.77\% & 79.98\% & 75.76\% & 93.36\% & 94.56\% & 87.89\% & 90.92\% & 93.92\% & 96.80\% & \textbf{93.75}\% \\ \hline
\end{tabular}
\label{tab:logic-syntax-rules}
\vspace{-0.7cm}
\end{table*}

\subsection{Concrete Requests: Comprehensive Evaluation}

\begin{sloppypar}
    To assess the effectiveness of our framework and the quality of the outputs it generates, we conduct experiments on concrete requests. These requests are expressed in natural language that can be straightforwardly transformed into command sequences for back-end layout tools. The concrete requests we crafted include detailed scenarios and explicit command directives designed to emulate interactions that designers might have with LayoutCopilot. \chg{Given the limited accessibility of real IC design data, we choose synthesized data for large-scale testing. To provide a clearer understanding of the types of requests generated, examples are provided below that represent typical testing cases used in our evaluation.} 
    \vspace{-0.2cm}
    \begin{tcolorbox}[colback=gray!5!white, colframe=gray!75!black, title=Example Requests for Testing Case, fonttitle=\bfseries, boxrule=0.75pt, arc=2mm, left=3pt, right=3pt, top=3pt, bottom=3pt]
        \scriptsize
        \textbf{Request 1:} \
        I have received a partially completed analog layout design, and I need to make some adjustments to improve the overall performance and routing efficiency. First, I would like to add symmetry constraints between devices M1 and M3, as well as M5 and M7, since they are part of a differential pair and should be laid out symmetrically. Next, I need to swap the positions of devices M11 and M13, as the current placement is causing significant routing congestion in that area of the layout. After making these initial placement changes, I will run the \texttt{route} command to see the updated routing results.
        
        \vspace{0.5em}
        \textbf{Request 2:} \
        Looking at the routing, I notice that net2 and net4 are critical signals that need to be as closely matched in length as possible. To achieve this, I will use the \texttt{netTopology} command to add guide points for net2 at coordinates (4500, 6800) and for net4 at (4800, 6800), so that the router can try to route these nets in a more symmetric fashion. In addition, I have identified a section of parallel alignment between net9 and net11 that is too close for comfort. I will use the \texttt{wwSpacing} command to increase the spacing between wire3 of net9 and wire5 of net11 to 150 units in the horizontal direction, to ensure proper signal integrity.
    \end{tcolorbox}
    \vspace{-0.2cm}
    We utilize GPT-3.5~\cite{LanguageModelsareFewShotLearners} and Claude-3~\cite{claude3} to generate a total of 1,250 cases, consisting of 1,134 valid concrete requests and 116 invalid requests. The invalid requests typically lack essential parameters or described operations in an order that violates the predefined rules in the manual of the layout tools, and we have verified these manually to confirm their invalidity. Each valid request can be completed by between 5 and 40 commands in Table~\ref{tab:command_set}, ensuring a variety of scenarios in requests. LayoutCopilot processes each concrete request to generate a command sequence that the layout tool could execute to fulfill the requirement. To fully validate the performance of LayoutCopilot, we performed a sanity check and functionality check on those test cases as described below:
\end{sloppypar}

\subsubsection{Sanity Check}\label{sec:sanity_check}

\begin{sloppypar}
    To validate the effectiveness of our agent configuration, we execute experiments under two distinct settings: one with the comprehensive configuration of the \chg{LLMs} through prompt engineering as outlined in Section \ref{subsec: agent configuration}, and the other by simply providing the \chg{LLMs} with the layout tool manual along with a straightforward task description that included input-output requirements. Additionally, comparative experiments are conducted using a single-agent setup under instructed conditions to demonstrate the necessity of the multi-agent methodology. In the multi-agent scenario, LayoutCopilot acts as a multi-agent collaborative framework, employing its Concrete Request Processor to manage requests from the test set. Conversely, the single-agent configuration merges all agent prompts mentioned in Section \ref{sec:Algorithm} into a single LLM agent aimed at encompassing the entire functionality of LayoutCopilot for handling the test requests. To minimize the impact of different LLM engines on the experimental outcomes, bulk testing is performed using GPT-3.5, GPT-4, and Claude-3 as the LLM engines.
\end{sloppypar}

\begin{sloppypar}
    Regarding testing standards, the sanity check covers formatting, validity, syntax, logic rules, and overall accuracy. (i) For the output format, the agent is required to generate processing status and results in JSON following regular text dialogue, ensuring that layout tools can accurately interpret the content and implement the necessary layout adjustments. (ii) Validity is measured by the rate at which the system correctly identifies and responds to invalid inputs. We expect LayoutCopilot to be capable of detecting invalid inputs and providing feedback to designers, such as commands missing specific parameters. (iii) \& (iv) We conduct syntactic and logical verification. Based on the command set in Table \ref{tab:command_set}, we establish four syntactic and two logical rules to ensure the integrity and logic of the command sequence. For example, a syntax rule ensures each command contains the correct number of parameters and one of the logic rules says that a device cannot appear in multiple symmetry pairs. Due to the page limit, we do not list all the rules here. (v) Successful processing of a request that either generates correctly formatted code adhering to these rules or accurately identifies an illegal request is considered correct handling and included in the overall accuracy statistics.
\end{sloppypar}


\begin{sloppypar}
    We conduct experiments on the previously mentioned test set of 1,250 cases following the outlined criteria, with the results displayed in Table \ref{tab:logic-syntax-rules}. Firstly, it is evident that instructions significantly enhance accuracy across various LLM engines, with average improvements of 6.49\% in logical verification, 4.85\% in syntactic verification, and an overall increase of 5.89\% in overall accuracy. This confirms the effectiveness of our LLM agent design described in Subsection \ref{subsec: agent configuration}. The case of single-agent without instruction is not listed because the correctness rate is too low to be informative, while the comparison of multi-agent with/without instruction is sufficient to illustrate the above conclusion. Secondly, the accuracy rates of the single-agent system are consistently lower than those for the multi-agent system, regardless of whether instructions and which LLM engines are used. This demonstrates the effectiveness of introducing a multi-agent methodology in this application scenario. \chg{Furthermore, the observed slight decrease in syntax performance for Claude-3 with instructions likely reflects a performance balancing: while instructions may slightly reduce performance on tasks already handled well, they improve weaker areas, resulting in better overall performance. Without instructions, certain tasks may perform better, but weaker areas can lower the overall performance. Thus, the value of instructions lies in their ability to enhance the overall performance balance.} Lastly, under conditions using Claude-3, multi-agent configuration, and with instructions, the overall correctness rate of the sanity check reaches 96.80\%. This confirms that LayoutCopilot's comprehensive capabilities are sufficient to meet practical application requirements. 
\end{sloppypar}


\subsubsection{Functionality Check}

\begin{figure}[tb]
    \centering
    \includegraphics[width=0.47\textwidth]{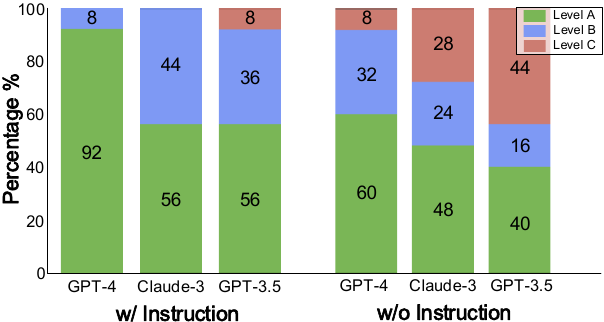}
    \caption{Functionality check with different LLM engines across instructional conditions.}
    \label{function_validation}
    \vspace{-0.8cm}
\end{figure}

\begin{sloppypar}
    To further assess the functional correctness of the outputs generated by LayoutCopilot on the test set, we randomly select 25(about 2\%) cases from the 1,250 generated results that have passed the sanity check. These are subjected to a manual functionality check to verify if they fully implemented the corresponding requests. Like the standards used in~\cite{he_chateda_2024}, we categorize the output results into three levels: A, B, and C. Level A corresponds to output functionally correct and displays clear analytical reasoning. Level B indicates outputs where the functionality has minor flaws, but the analytical reasoning is clear, allowing designers to identify and rectify issues easily. Level C represents outputs that fail both functionally and conceptually, offering little valuable information to help diagnose problems. In practical scenarios, the last thing we can accept is for LayoutCopilot to give us advice that is incorrect, unjustified, and difficult to fix. Thus, eliminating level C is the focus of our design efforts.
\end{sloppypar}

\begin{sloppypar}
    We conduct tests using the Concrete Request Processor both with and without instructions, corresponding to the multi-agent section in Table 2. The single-agent scenario is omitted due to its significantly lower pass rate in sanity check, which does not offer valuable comparative insights. As summarized in Figure~\ref{function_validation}, functionality check results with instructions significantly outperform those without instructions across different LLM engines: on one hand, Level C, which is not uncommon without instructions, almost disappears under instructed conditions, with no more than 8\% in GPT-3.5. This indicates that the use of instructions, as discussed in Section 3.3, substantially enhances the practicality of LayoutCopilot, where the majority of outputs are either correct or can be easily corrected through simple interactions. On the other hand, the proportion of Level A increases by 16\% to 32\%, reaching up to 92\% with GPT-4 at the highest. Additionally, the percentage of C level under instructed conditions is near zero regardless of the LLM engine used. This demonstrates that LayoutCopilot can get good results under conditions where various LLMs are used as engines, even if there are gaps in the capabilities of these LLM engines. Overall, the functionality validation demonstrates that LayoutCopilot's design significantly enhances the output quality, proving its practical utility.
\end{sloppypar}

\chg{
\subsubsection{Classification Accuracy Evaluation}
\begin{sloppypar}
To evaluate the accuracy of the Classifier Agent, we conducted experiments using GPT-3.5 and Claude-3. The test set consisted of 1000 Concrete Requests and 1000 Abstract Requests, generated following the method detailed in Section~\ref{sec:sanity_check} to maintain consistency across experiments. The results, as presented in Table~\ref{tab:classification_accuracy}, show that the Classifier Agent achieved high accuracy across both models. Using GPT-3.5, the agent correctly identified 976 out of 1000 concrete requests and 981 out of 1000 abstract requests. With Claude-3, both concrete and abstract request classifications reached 100\% accuracy. Such high accuracy may be attributed to the relative simplicity of the task and the clarity of the classification criteria, as well as the strong enough pre-trained LLMs. These results affirm the effectiveness and reliability of the Classifier Agent in accurately distinguishing between concrete and abstract requests, providing strong support for its role in the LayoutCopilot framework.
\end{sloppypar}
}

\begin{table}[tb]
  \caption{Classification Accuracy of the Classifier Agent \\on GPT-3.5 and Claude-3}
  \centering
  \resizebox{0.25\textwidth}{!}{%
  \begin{tabular}{|c|c|c|}
    \hline
    \multirow{2}{*}{\textbf{Request Type}} & \multicolumn{2}{c|}{\textbf{Accuracy (\%)}} \\ \cline{2-3}
    & \scriptsize{GPT-3.5} & \scriptsize{Claude-3} \\ 
    \hline
    Concrete & 97.6 & 100 \\ 
    Abstract & 98.1 & 100 \\ 
    \hline
  \end{tabular}%
  }
  \label{tab:classification_accuracy}
  \vspace{-0.7cm}
\end{table}



\subsection{Abstract Requests: Case Studies}
\begin{figure*}[tb]
    \centering
    \includegraphics[width=0.96\textwidth]{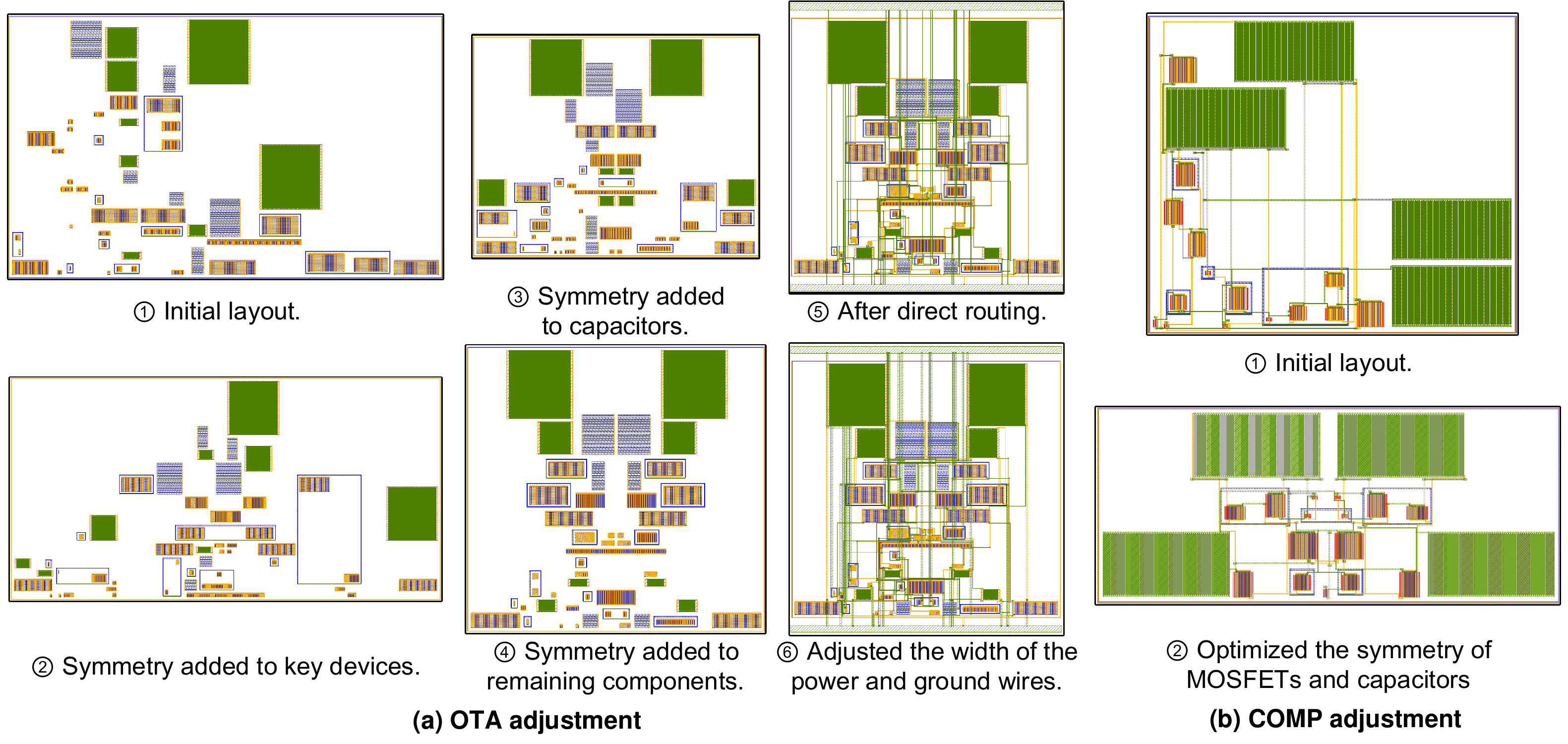}
    \caption{Interactive LLM-powered layout adjustment on OTA and COMP.}
    \label{LayoutResult}
    \vspace{-0.4cm}
\end{figure*}

\begin{figure}[tb]
    \centering
    \includegraphics[width=0.47\textwidth]{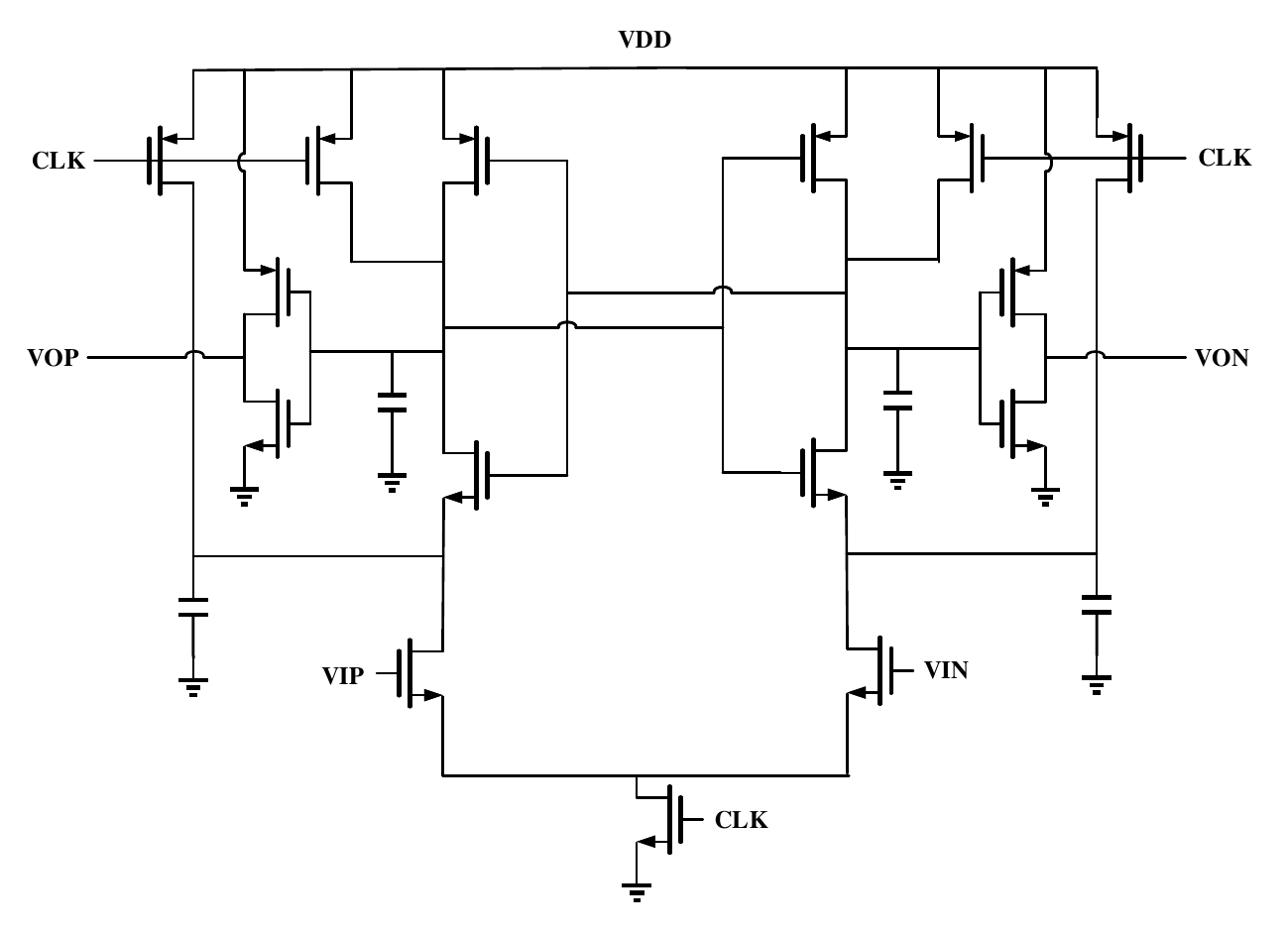}
    \caption{The circuit schematic of the COMP.}
    \label{COMP_schematic}
    \vspace{-0.8cm}
\end{figure}

\begin{sloppypar}
    In this subsection, we illustrate LayoutCopilot's capabilities in addressing abstract requests by optimizing layouts in real-world designs. The schematics of these two circuits are shown in Figure \ref{OTA_schematic} and Figure \ref{COMP_schematic}, and their layout adjustment processes are depicted in Figure \ref{LayoutResult} (a) and (b), respectively.
\end{sloppypar}

\subsubsection{Operational Transconductance Amplifier (OTA)}
\begin{figure}[tb]
    \centering
    \includegraphics[width=0.47\textwidth]{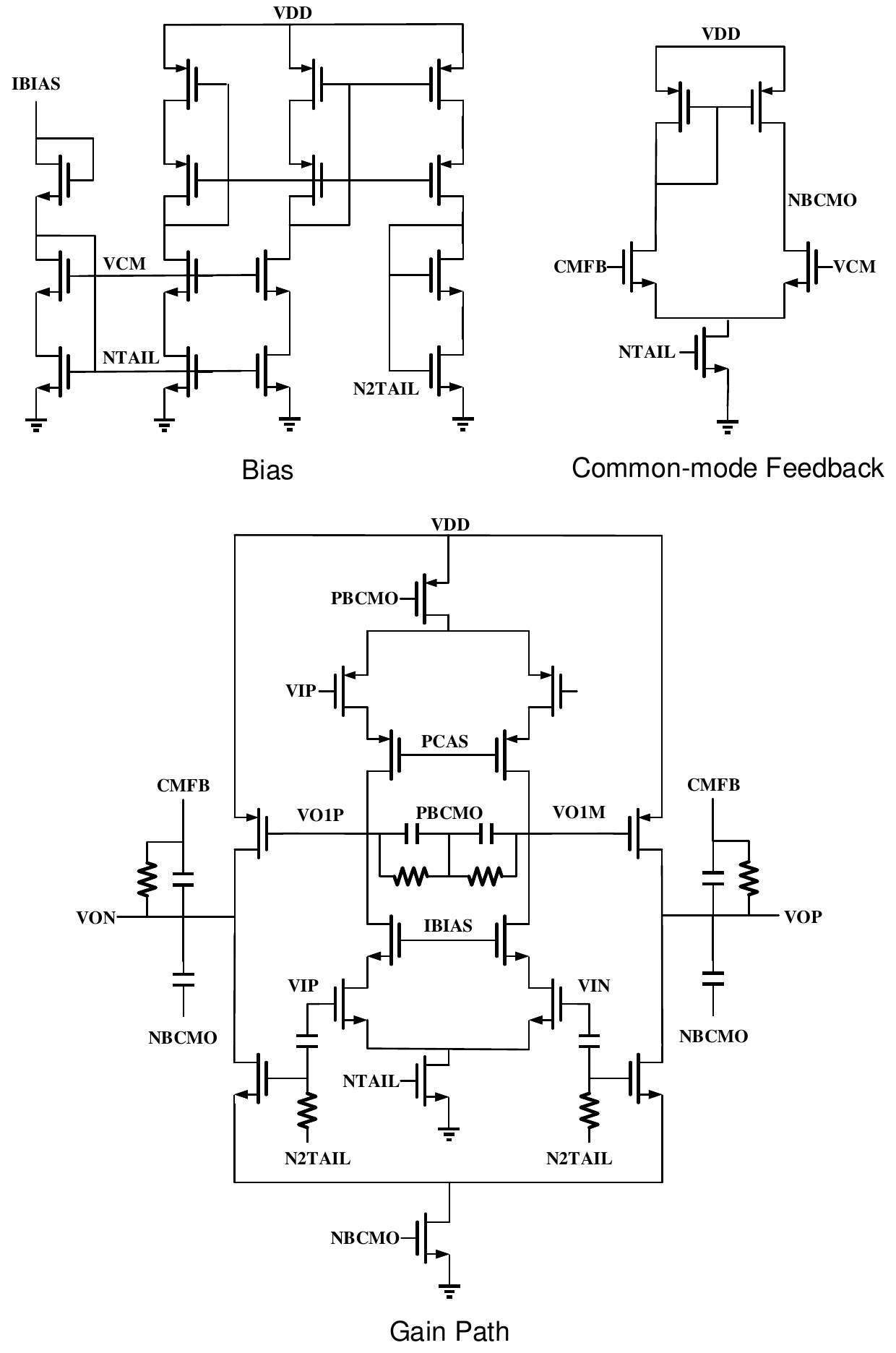}
    \caption{The circuit schematic of the OTA.}
    \label{OTA_schematic}
    \vspace{-0.6cm}
\end{figure}

\begin{sloppypar}
    Figure \ref{LayoutResult}(a) showcases the layout optimization process for an OTA, whose schematic is shown in Figure \ref{OTA_schematic}. We employ the MAGICAL~\cite{xu_magical_2019, chen_magical_2021, chenMAGICALOpenSource2021} to generate the initial placement layout \ding{172}, without any constraints applied. The initial layout is irrelevant to our framework, which focuses solely on modifications rather than origins. From layout \ding{172}, we engage LayoutCopilot to suggest improvements during the placement phase, simulating real-world scenarios faced by designers. LayoutCopilot proposes a series of high-level solutions, briefly shown as follows:
\end{sloppypar}

\begin{tcolorbox}[
    arc=0mm,outer arc=0mm,
    boxrule=0mm,toprule=0.5mm,bottomrule=0.5mm,left=1mm,right=1mm,
    titlerule=0mm,toptitle=0mm,bottomtitle=0mm,top=0mm,
    title = {\scriptsize\textbf{Designer:} \newline \scriptsize{I got an OTA layout, but it has poor performance. How can I improve it?}},
    fontupper=\small,
    colback=white,
    coltitle=black,
    colbacktitle=white!90!gray   
]
\scriptsize
\textbf{LayoutCopilot:}

Improving the performance of your OTA layout can involve...

Here are some potential high-level solutions:

\begin{enumerate}
\def\labelenumi{\arabic{enumi}.}
\item
  Enhance Symmetry with \texttt{symAdd} ...
\item
  Improve Matching with \texttt{deviceMove} and \texttt{deviceSwap} ...
\item
  Reduce Parasitics with \texttt{wireWidth} and \texttt{wireSpacing} ...
\item
  Prevent Crosstalk with \texttt{wireSpacing} ...
\item
  Improve Routing with \texttt{netPriority} and \texttt{netTopology} ...
\item
  ...
\end{enumerate}

\end{tcolorbox}











\begin{sloppypar}
    Following this, we select the `Enhance Symmetry' option and provide LayoutCopilot with the netlist, asking for concrete implementation steps. LayoutCopilot analyzes the netlist and identifies key circuit components that should have symmetry constraints applied:
\end{sloppypar}

\vspace{-0.3cm}
\begin{tcolorbox}[
    arc=0mm,outer arc=0mm,
    boxrule=0mm,toprule=0.5mm,bottomrule=0.5mm,left=1mm,right=1mm,
    titlerule=0mm,toptitle=0mm,bottomtitle=0mm,top=0mm,
    title = {
    \scriptsize
    \textbf{Designer:} \newline \scriptsize{Okay, I prefer to start with the Enhance Symmetry optimization since the current symmetry looks terrible. Here is the circuit netlist: [Netlist]. \newline Which devices should I add symmetry to? Please help me with the corresponding commands. }
    },
    fontupper=\small,
    colback=white,
    coltitle=black,
    colbacktitle=white!90!gray   
]
\scriptsize
\textbf{LayoutCopilot:}

Let\textquotesingle s proceed with analyzing the netlist to identify
critical devices...

\emph{Devices for Symmetry:}

\begin{enumerate}
\def\labelenumi{\arabic{enumi}.}
\item
  Differential Input Pair: M34 (net0130 VIM PTAIL VDD) and M35 (net0132
  VIP PTAIL VDD) handle the input signals ...
\item
  Output Stage: M71 (VOP VO1M VDD VDD) and M70 (VOM VO1P VDD
  VDD) are used in the output stage ...
\item
  Tail Current Source: M1 (NTAIL NBIAS\_TAIL GND GND) functions as a
  tail current source...
\item
  ...
\end{enumerate}

\emph{Command Sequence:}

\begin{verbatim}
symAdd M34 M35  # Differential input stage.
symAdd M71 M70  # Output stage.
symAdd M1 M1    # Self-symmetry for tail.
...
\end{verbatim}
% ...
\end{tcolorbox}
\vspace{-0.3cm}










% ...

\begin{sloppypar}
    After executing these commands, the resultant layout is shown in \ding{173}. Observing layout \ding{173}, we note that large capacitors were not symmetrically constrained, leading to a loose and cluttered arrangement. Upon receiving more concrete requests from us, LayoutCopilot applies symmetry constraints to all capacitors based on sizing results and connectivity, resulting in layout \ding{174}. For the sake of simplicity, the subsequent replies are not shown specifically.
\end{sloppypar}


\begin{sloppypar}
    Further, we direct LayoutCopilot to apply symmetry to all possible components, including resistors that were previously unconsidered. After identifying suitable components in the netlist, LayoutCopilot generates and executes the corresponding commands, yielding layout \ding{175}. With the placement appearing optimally refined visually, we proceed with a basic routing command to produce layout \ding{176}. Subsequent post-layout simulation reveals excessive parasitics, leading us to seek further solutions from LayoutCopilot. To address these issues, we engaged in a detailed examination of the critical nets and implemented priority routing strategies as follows:
\end{sloppypar}
\vspace{-0.2cm}
\begin{tcolorbox}[
    arc=0mm,outer arc=0mm,
    boxrule=0mm,toprule=0.5mm,bottomrule=0.5mm,left=1mm,right=1mm,
    titlerule=0mm,toptitle=0mm,bottomtitle=0mm,top=0mm,
    title = {\scriptsize\textbf{Designer:} \newline \scriptsize{After symmetry, let's have some improvements on Routing as well, I would like to give higher priority to the critical nets to reduce their wire lengths and parasitics. \newline Please identify them and help me with the corresponding commands.  }},
    fontupper=\small,
    colback=white,
    coltitle=black,
    colbacktitle=white!90!gray   
]
\scriptsize
\textbf{LayoutCopilot:}

Let\textquotesingle s proceed with analyzing the netlist to identify critical nets...

\emph{Critical Nets:}

\begin{enumerate}
\def\labelenumi{\arabic{enumi}.}
\item
  Input Differential Pair Connections: Gate and drain of M34 (net0130 VIM PTAIL VDD) and M35 (net0132 VIP PTAIL VDD) ...
\item
  Output Differential Pair Connections: Gate and drain of M71 (VOP VO1M VDD VDD) and M70 (VOM VO1P VDD VDD) ...
\item
  Input Stage Compensation Capacitor Connections: Connections of C3 (VIM net096 GND) and C2 (VIP net092 GND) ...
\item
  Load Compensation Capacitor Connections: ...
\end{enumerate}

\emph{Command Sequence:}

\begin{verbatim}
netPriority net0130 10  # Drain of M34.
netPriority VIM 10      # Gate of M34.
netPriority net0132 10  # Drain of M35.
netPriority VIP 10      # Gate of M35.
...
netPriority net096 8    # Connection of C3.
netPriority net092 8    # Connection of C2.
...
\end{verbatim}
% ...
\end{tcolorbox}
\vspace{-0.3cm}

\begin{sloppypar}
    Subsequent to modifying the routing priority as per the above commands, we also increased the wire widths of power and ground lines (not elaborated here due to space constraints), finally achieving the results shown in layout \ding{177}. With the post-layout results now meeting our specifications as shown in Table \ref{tab:OTA&CMP}, we terminated the iterative optimization process.
\end{sloppypar}

\begin{sloppypar}
    We conduct the optimization process under TSMC 40nm technology to verify the circuit performance after modifications, utilizing Cadence Virtuoso and Mentor Graphics Calibre for post-layout simulation. As we progressed through the interactive layout adjustment process with LayoutCopilot, the layout area initially increased during the placement phase but eventually reduced to 66\% of its original size, as depicted in Table \ref{tab:benchmark area statics} and Figure \ref{LayoutResult}. Given the complex relationship between placement \& routing in analog circuits and their post-layout performance, improvements do not follow a linear progression from \ding{172} to \ding{176},  as described earlier for changes in the layout area. For simplicity, we compare the post-layout performance results between the initial layouts (without constraints) and the final layouts adjusted through interaction with LayoutCopilot, summarized in Table \ref{tab:OTA&CMP}.
\end{sloppypar}

\begin{sloppypar}
    In the initial layouts, due to excessive parasitics, the Gain was negative, the unity-gain bandwidth (UGB) and phase margin (PM) are substantially poor, and the common-mode rejection ratio (CMRR) is significantly low, showing a large discrepancy with the pre-layout schematic results which do not account for parasitic extractions. After adjustments made using LayoutCopilot, while there remains a disparity in UGB compared to the schematic results, both Gain and PM have closely approached the schematic levels, with CMRR showing substantial improvement over the initial layout. These results underscore the efficacy of LayoutCopilot in facilitating layout optimization through natural language interaction and reducing both the learning and coding time for designers, demonstrating its potential to serve as a powerful assistant for analog layout designers.
\end{sloppypar}


\begin{table}[tb]
  \caption{{Placement area statistics.}}
  \label{tab:benchmark area statics}
  \centering
  \resizebox{.35\textwidth}{!}{
  \begin{tabular}{|c|c|c|c|}
  \toprule
  \multirow{2}{*}{\textbf{Benchmark}}  & \textbf{Technology}  & \textbf{Placement} & \textbf{Area} \\
  \multicolumn{1}{|c|}{}  & \textbf{Node} & \multicolumn{1}{c|}{\textbf{Area}} & \multicolumn{1}{|c|}{\textbf{Ratio}}\\ 
  \midrule
  OTA-1  & \multirow{4}{*}{TSMC40} & $ 83.4 \times 124.2\mu m^{2} $ & 1.00 \\
  OTA-2  & \multicolumn{1}{c|}{} & $ 85.0 \times 148.9\mu m^{2} $ & 1.22\\ 
  OTA-3  & \multicolumn{1}{c|}{} & $ 80.6 \times 94.3 \mu m^{2} $ & 0.73\\
  OTA-4  & \multicolumn{1}{c|}{} & \textbf{$ 85.4 \times 80.4 \mu m^{2} $}  & {0.66} \\
  \midrule
  COMP-1 & \multirow{2}{*}{TSMC28} & $ 38.3 \times 39.7\mu m^{2} $ & 1.00 \\
  COMP-2  & \multicolumn{1}{c|}{} & \textbf{$ 24.0 \times 46.9\mu m^{2} $} & {0.74}\\ 
  \bottomrule
  
  \end{tabular}
  }
\end{table}

 \begin{table}[tb]
   \caption{{Comparison between LayoutCopilot and MAGICAL~\cite{xu_magical_2019, chen_magical_2021, chenMAGICALOpenSource2021} without constraints.}}
  \label{tab:OTA&CMP}
  \resizebox{.48\textwidth}{!}{
  \begin{tabular}{|c|c|c|c|c|}
  \hline
  \toprule
    \multicolumn{2}{|c|}{\multirow{2}{*}{\textbf{Benchmark}}} & \multirow{2}{*}{\textbf{Schematic}} & \textbf{MAGICAL~\cite{xu_magical_2019, chen_magical_2021, chenMAGICALOpenSource2021}}   & \multirow{2}{*}{\textbf{LayoutCopilot}} \\
    \multicolumn{2}{|c|}{}& & \textbf{w/o Constraints} & \\
  \midrule
    \multirow{4}{*}{OTA} & Gain (dB) & 38.63  & -8.75 & \textbf{38.26}  \\
    \multicolumn{1}{|c|}{}& UGB (MHz) & 6.85  & -- & \textbf{4.42}  \\
    \multicolumn{1}{|c|}{} & CMRR (dB) & --   & 27.3 & \textbf{58.7}  \\
    \multicolumn{1}{|c|}{} & PM (degree) & 70.98  & -- & \textbf{76.28}  \\
  
  \hline

  \multirow{4}{*}{COMP}  & CMP\_Delay (ns) & 3.3 & 6.3  & 6.7   \\
  \multicolumn{1}{|c|}{} & Noise (uV) & 50.3 & 30.9  & \textbf{13.9}  \\
  \multicolumn{1}{|c|}{} & RST\_Delay (ps) & 89.8 & 165.8 & 538.7  \\ 
  \multicolumn{1}{|c|}{} & Power (uW) & 19.9 & 32.0 & \textbf{31.71}  \\
  \bottomrule
  \hline
  \end{tabular}
  }
  \vspace{-0.7cm}
\end{table}

\subsubsection{Comparator (COMP) }
\begin{sloppypar}
    A similar process for the COMP adjustments is employed, now utilizing the TSMC 28nm technology as depicted in Figure \ref{LayoutResult} (b). The schematic of the COMP is shown in Figure \ref{COMP_schematic}. We optimize the symmetry for transistors and capacitors through interactions with LayoutCopilot, transforming an initial layout in \ding{172} into the improved layout in \ding{173}. As summarized in the lower half of Table \ref{tab:OTA&CMP}, the post-layout simulation result shows that we have significantly improved noise performance at the cost of delay and enhanced power efficiency compared to the initial layout. This adjustment process for both the OTA and CMP exemplifies how LayoutCopilot leverages its comprehensive knowledge base to provide actionable recommendations, effectively optimizing the layout of circuits with varying topologies and improving their post-simulation performance.
\end{sloppypar}







%% file: doc/conclu.tex
\section{Conclusion}
\label{sec:Conclusion}


In this work, we propose a multi-agent collaborative framework powered by LLMs for interactive analog layout design. LayoutCopilot can not only convert natural language instructions into executable script commands but also interpret high-level design intents into actionable suggestions for implementation. Technically, LayoutCopilot employs a multi-agent methodology alongside prompt engineering. Validated in both TSMC28 and TSMC40, the experimental results demonstrate the robustness and benefits of LayoutCopilot. Specifically, LayoutCopilot achieves high accuracy in handling concrete requests for layout adjustments and effectively improves layout performance in addressing abstract layout optimization requests. We believe this work can provide new insights into solving usability issues of interactive EDA tools and facilitate the development of automation tools for analog circuits.

%% file: doc/bio.tex
\vspace{-1.5cm}
\begin{IEEEbiography}[{\includegraphics[width=1in,height=1.25in,clip,keepaspectratio]{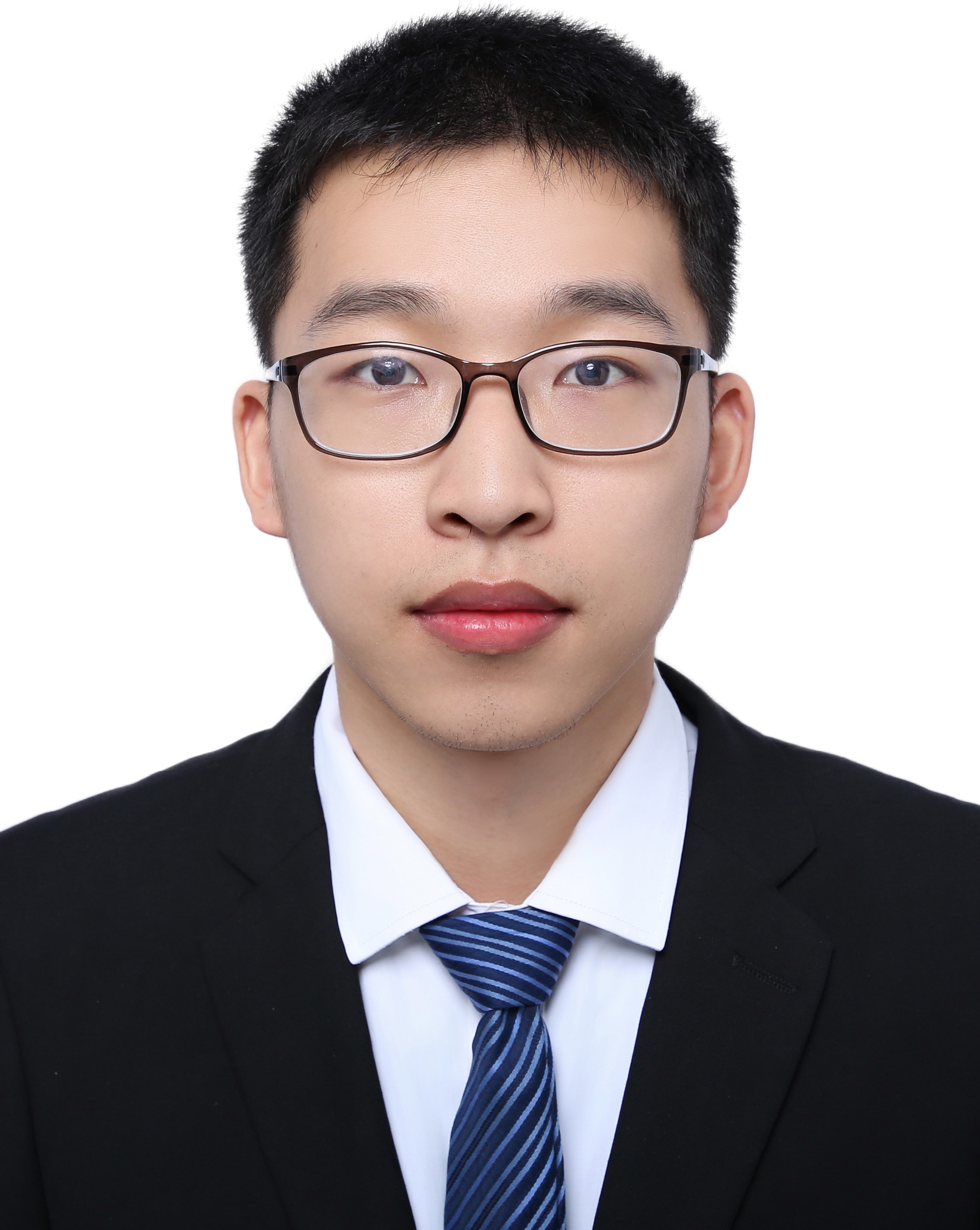}}]{Bingyang Liu}
is currently an undergraduate student at the School of Electronics Engineering and Computer Science, Peking University, Beijing, China. His research interests include analog design automation and ML for EDA.
\end{IEEEbiography}

\vspace{-1.5cm}
\begin{IEEEbiography}[{\includegraphics[width=1in,height=1.25in,clip,keepaspectratio]{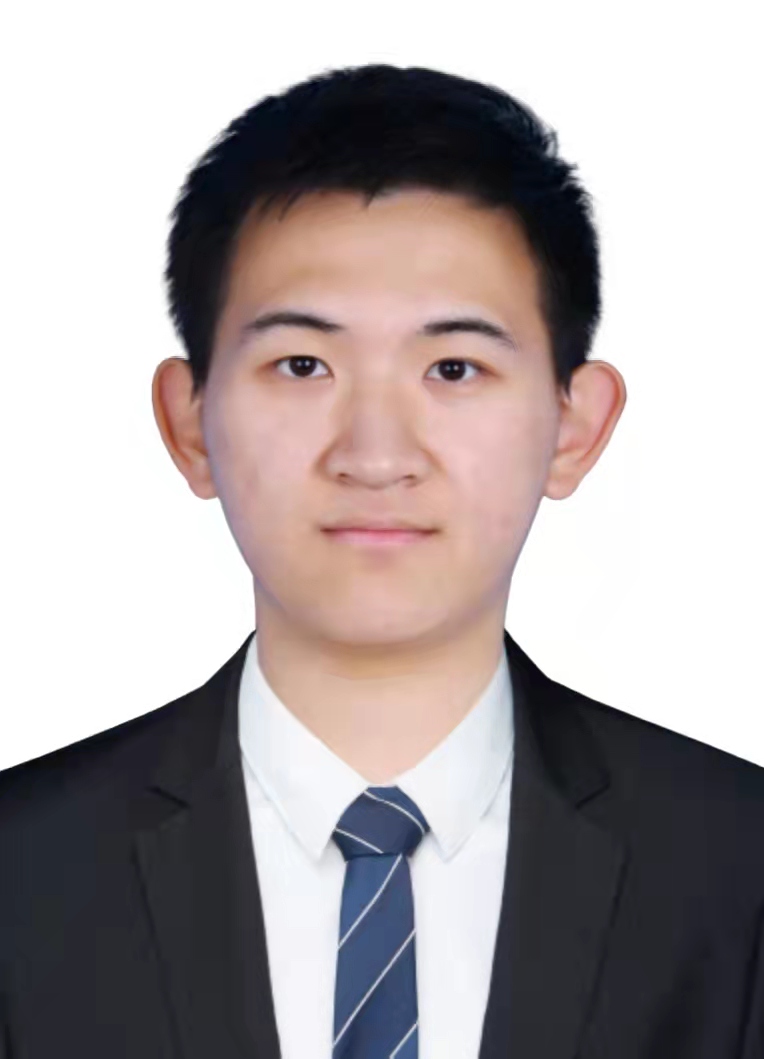}}]{Haoyi Zhang}
received the B.S. degree in Microelectronics from BeiHang University in 2022. He is currently pursuing the Ph.D degree in microelectronics with Peking University, Beijing, China.
His research interest inclues analog design automation, mixed-signal circuit design.
\end{IEEEbiography}

\vspace{-1.5cm}
\begin{IEEEbiography}[{\includegraphics[width=1in,height=1.25in,clip,keepaspectratio]{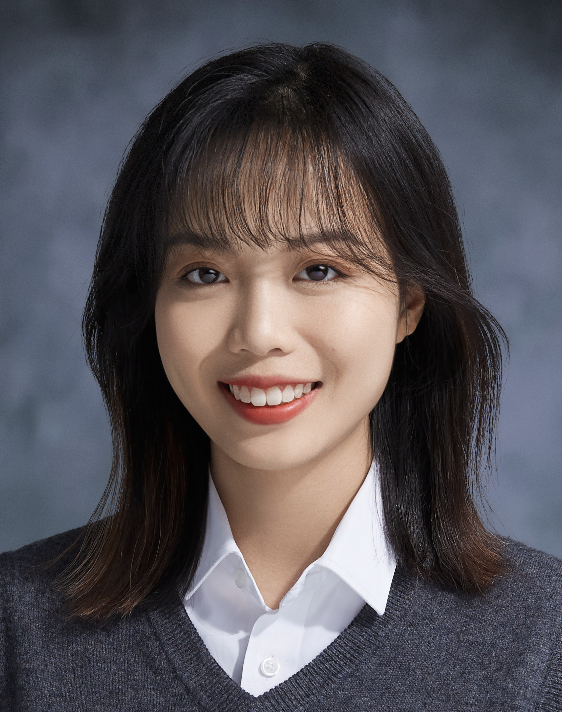}}]{Xiaohan Gao}
received the B.S. degree in Computer Science from Peking University in 2020.
She currently is a Ph.D. student, advised by Prof. Yibo Lin, with the School of Computer Science at Peking University.
Her research interests include layout design automation for analog and mixed-signal circuits, incorporating machine learning techniques with physical design and design manufacturability, and abstracting description from layout designs.
\end{IEEEbiography}

\vspace{-1.5cm}
\begin{IEEEbiography}[{\includegraphics[width=1in,height=1.25in,clip,keepaspectratio]{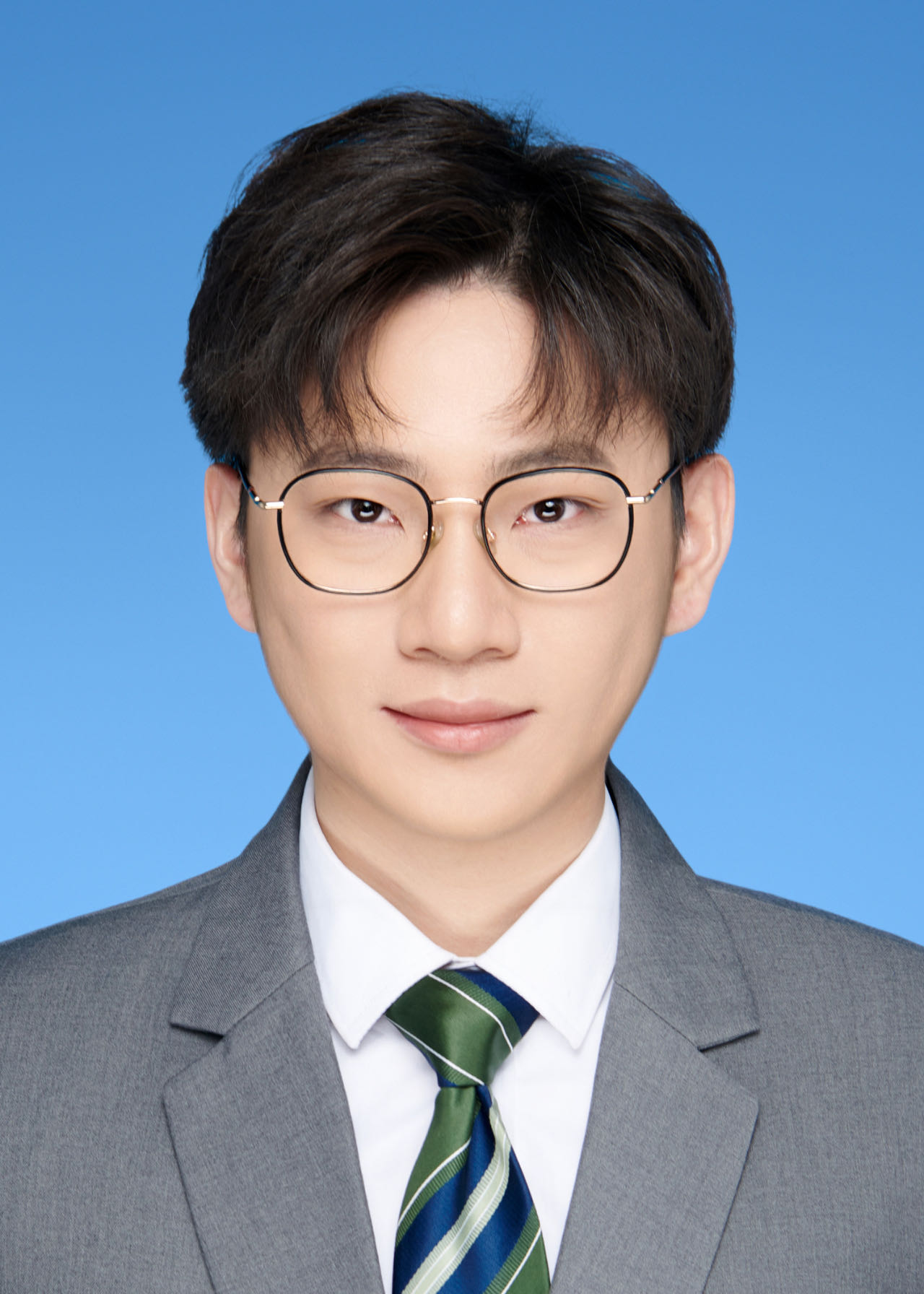}}]{Zichen Kong}
received the B.S. degree from the School of Electronics Engineering and Computer Science at Peking University, Beijing, China, in 2023. He is currently pursuing the Ph.D. degree with the School of Integrated Circuits, Peking University, Beijing, China. His research interests include CMOS image sensor and analog design automation.
\end{IEEEbiography}

\vspace{-1.5cm}
\begin{IEEEbiography}[{\includegraphics[width=1in,height=1.25in,clip,keepaspectratio]{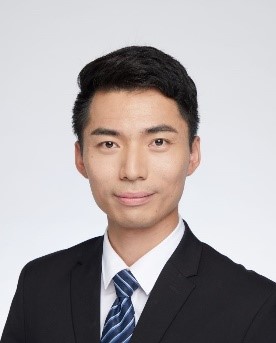}}]{Xiyuan Tang}
(S’17 -M’19-SM’24) received the B.Sc. degree (Hons.) from the School of Microelectronics, Shanghai Jiao Tong University, Shanghai, China, in 2012, and the M.S. and Ph.D. degree in electrical engineering from The University of Texas at Austin, Austin, TX, USA, in 2014 and 2019 respectively.

He is currently an Assistant Professor at Peking University, Beijing, China. He was a Design Engineer with Silicon Laboratories, Austin, TX from 2014 to 2017, where he was involved in the RF receiver design. From 2019-2021, he was a postdoctoral researcher at the University of Texas at Austin, Austin, TX. His research interests include digitally assisted data converters, low-power mixed-signal circuits, and analog data processing. He 

Dr. Tang serves on the Technical Program Committees (TPC) of ISSCC. He also serves as an associate editor for IEEE Solid-State Circuits Letters. He was a recipient of IEEE Solid-State Circuits Society Rising Stars in 2020, Best Paper Award at Silicon Labs Tech Symposium in 2016, National Scholarship in 2011, and Shanghai Scholarship in 2010.
\end{IEEEbiography}

\vspace{-1.5cm}
\begin{IEEEbiography}[{\includegraphics[width=1in,height=1.25in,clip,keepaspectratio]{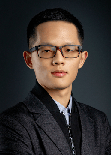}}]{Yibo Lin}
(M'19) received the B.S. degree in Microelectronics from Shanghai Jiaotong University in 2013.
He obtained his Ph.D. degree in Electrical and Computer Engineering from the University of Texas at Austin in 2018 advised by Prof. David Z. Pan.
He worked as a postdoctoral researcher at the University of Texas at Austin from 2018 to 2019.
He currently is an assistant professor in the School of Integrated Circuits at Peking University.
His research interests include physical design, machine learning applications, and heterogeneous computing in VLSI CAD.
He is a recipient of the Best Paper Awards at preimier EDA/CAD journals/conferences like TCAD, DAC, DATE, ISPD, etc.
\end{IEEEbiography}

\vspace{-1.5cm}
\begin{IEEEbiography}[{\includegraphics[width=1in,height=1.25in,clip,keepaspectratio]{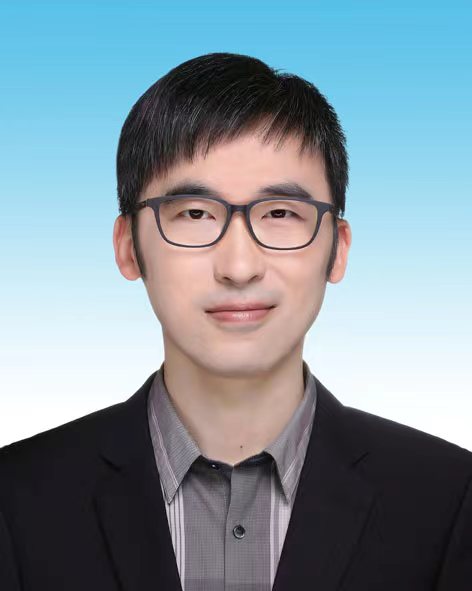}}]{Runsheng Wang}
(S’07–M’11)
received the B.S. and Ph.D. (highest honors) degrees from Peking University, Beijing, China, in 2005 and 2010, respectively. 

From November 2008 to August 2009, he was a Visiting Scholar with Purdue University, West Lafayette, IN, USA. He joined Peking University in 2010, where he is currently a Professor at the School of Integrated Circuits and is serving as the Associate Dean of the School of EECS. He has authored/coauthored 1 book, 4 book chapters, and about 200 scientific papers, including more than 40 papers published in \textit{International Electron Devices Meeting} (IEDM) and \textit{Symposium on VLSI Technology} (VLSI-T). He has been granted 19 US patents and 38 Chinese patents. His current research interests include nanoscale CMOS devices and reliability, design automation, and new-paradigm computing.

Dr. Wang was awarded the IEEE EDS Early Career Award by the IEEE Electron Device Society (EDS), National Distinguished Young Scholars by the National Natural Science Foundation of China (NSFC), Natural Science Award (First Prize) by the Ministry of Education (MOE) of China, and many other awards. He serves on the Editorial Board of \textit{IEEE TRANSACTIONS ON ELECTRON DEVICES}, and \textit{SCIENCE CHINA: Information Sciences}, and has served on the Technical Program Committee of many IEEE conferences, including \textit{IEDM}, \textit{IRPS}, etc.
\end{IEEEbiography}

\begin{IEEEbiography}[{\includegraphics[width=1in,height=1.25in,clip,keepaspectratio]{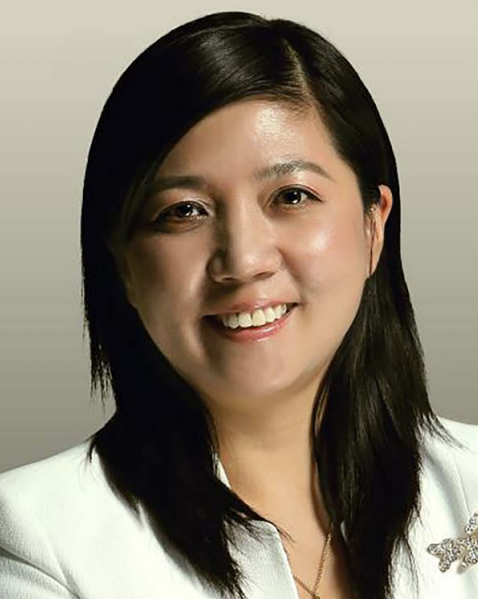}}]{Ru Huang}
(Fellow, IEEE) received the B.S. (Hons.) and M.S. degrees in electronic engineering from Southeast University, Nanjing, China, in 1991 and 1994, respectively, and the Ph.D. degree in microelectronics from Peking University, Beijing, China, in 1997.
Since 1997, she has been a faculty member with Peking University, where she is currently a Boya Chair Professor. She is an elected Academician of the Chinese Academy of Science, an elected member of TWAS Fellow. She has authored or coauthored five books, five book chapters, and more than 300 papers, including more than 100 papers in IEDM (46 IEDM papers from 2007 to 2021), VLSI Technology Symposium, IEEE EDL, and IEEE T-ED. She has delivered over 50 keynote/invited talks at conferences and seminars. She has been granted over 300 patents including 49 U.S. patents. Her research interests include nano-scaled CMOS devices, ultra-low-power new devices, new device for neuromorphic computing, emerging memory technology, and device variability/reliability.
\end{IEEEbiography}